\documentclass[reqno,11pt]{amsart}
\usepackage{amssymb,amsthm}

\usepackage{a4}
\numberwithin{equation}{section}

\usepackage[mathscr]{eucal}

\newcounter{mnotecount}[section]
\renewcommand{\themnotecount}{\thesection.\arabic{mnotecount}}

\newcounter{mymnotecount}[section]
\renewcommand{\themymnotecount}{\thesection.\arabic{mymnotecount}}
\newcommand{\mymnote}[1]
{\protect{\stepcounter{mymnotecount}}$^{\mbox{\footnotesize $%\!\!\!\!\!\!\,
\bullet$\themnotecount}}$ \marginpar{%\color{red}%
\raggedright\tiny\em $\!\!\!\!\!\!\,\bullet$\themymnotecount: #1} }
\renewcommand{\mymnote}[1]{}

\renewcommand{\Re}{\mathbb R}
\newcommand{\half}{\frac{1}{2}}

\newcommand{\eps}{\epsilon}
\newcommand{\Bo}{\mathcal B}
\newcommand{\Sp}{\mathcal S}
\newcommand{\id}{\mathbf{id}}
\newcommand{\TT}{\mathcal T}
\newcommand{\MM}{\mathcal M}
\newcommand{\LL}{\mathcal L}
\newcommand{\hphi}{\widehat \phi}
\newcommand{\FF}{\mathcal F}

\newcommand{\bU}{\bar U}
\newcommand{\ii}{\mathbf i}

\newcommand{\Proj}{{\mathbb P}}
\newcommand{\BProj}{\Proj_{\Bo}}
\newcommand{\fProj}{\Proj_{f^{-1}(\Bo)}}

\newcommand{\fProjN}{\fProj}
\newcommand{\fIdN}{\fId}

\newcommand{\Id}{\mathbb I}
\newcommand{\fId}{{\Id}_{f^{-1}(\Bo)}}

\newcommand{\QQ}{\mathbb Q}

\newcommand{\XX}{\mathbb X}
\newcommand{\YY}{\mathbb Y}

\newcommand{\ZZ}{\mathcal Z}

\newcommand{\Lie}{\mathcal L}

\newcommand{\oleq}{\lesssim}

\newcommand{\dHodge}{\operatorname{d}}

\theoremstyle{plain}
\newtheorem{thm}{Theorem}[section]
\newtheorem{cor}[thm]{Corollary}
\newtheorem{lemma}[thm]{Lemma}

\newtheorem{prop}[thm]{Proposition}
\newtheorem{remark}{Remark}[section]

\title{Rotating elastic bodies in Einstein gravity}
\author[L. Andersson]{Lars Andersson${}^\dagger$} \email{larsa@math.miami.edu}
\thanks{${}^\dagger$ Supported in part by the NSF, under
contracts no. DMS 0407732 and DMS 0707306 with the University of Miami.}
\address{Albert Einstein Institute, Am M\"uhlenberg 1, D-14476 Potsdam,
  Germany \and
Department of Mathematics, University of Miami, Coral Gables, FL
33124, USA}

\author[R. Beig]{Robert Beig${}^{\ddagger}$
} \email{beig@ap.univie.ac.at}
\address{Gravitational Physics, Faculty of Physics, University of
Vienna,
\newline
Boltzmanngasse 5, A-1090 Vienna, Austria}
\thanks{${}^{\ddagger}$ Supported in part by Fonds zur F\"orderung der
Wissenschaftlichen Forschung project no. P20414-N16.}
\author[B. Schmidt]{Bernd G. Schmidt} \email{bernd@aei.mpg.de}
\address{Max-Planck-Institut f\"ur Gravitationsphysik,
Albert-Einstein-Institut,
\newline 
Am M\"uhlenberg 1, D-14476 Golm, Germany}

\begin{document}
\date{November 6, 2008}

\begin{abstract}
We prove that, given a stress-free, axially symmetric elastic body,
there exists, for sufficiently small values of the gravitational
constant and of the angular frequency, a unique stationary
axisymmetric solution to the Einstein equations coupled to the
equations of relativistic elasticity with the body 
performing rigid rotations 
around the symmetry axis at the given angular frequency.
\end{abstract}

\maketitle

\section{Introduction} \label{sec:intro}
In the paper \cite{ABS}, we constructed for the first time static,
self-gravitat\-ing elastic bodies in general relativity with no symmetries.
Here we build on the
ideas and techniques introduced in that paper
to construct
solutions to the Einstein equations
describing steady states of self-gravitating matter
in rigid rotation.
The
matter model we use is, as in \cite{ABS},
that of a perfectly elastic solid. We make the minimal symmetry assumptions
necessary for a steady state in rigid rotation,
namely we assume that the reference
body has an axis of symmetry. Further, we assume that the elastic material is
isotropic. This condition, which was not needed in the static case, is
necessary for our construction in the case of a rotating body.

The
only class of solutions to the stationary Einstein equations
with rotating matter previously known are the rotating perfect
fluid solutions constructed by Heilig \cite{heilig}  for a certain class of
equations of state.
In the Newtonian theory, existence of
steady states of self-gravitating perfect fluids in rigid
rotation was
established by Lichtenstein, see \cite{heilig:newtonian} for a modern
presentation, and by Beig and Schmidt \cite{beig:schmidt:rotating}
for the case of elastic matter.
All these solutions, including the ones constructed in the present
paper are in addition to being stationary, also axisymmetric.

In the Newtonian theory, two families of non-axisymmetric rotating fluid configurations in
explicit form are
known,
see \cite{lindblom:1992} and references therein.
These families of solutions are the Dedekind ellipsoids,
and the Jacobi ellipsoids, which in the
language of general relativity have helical, but no stationary or
axial symmetry.

One expects asymptotically flat rotating solutions of the Einstein equations,
which are not axially symmetric to be radiating, and hence non-stationary.
However, if one relaxes the condition that asymptotic flatness holds
in the usual
sense, it may be possible to construct helically symmetric
solutions of the Einstein equations which are not axially symmetric.
See
\cite{Beig:2006pr} for a study helically symmetric solutions in the special
relativistic case.
An argument to the effect that
axisymmetry necessarily holds for a rotating fluid in general relativity
was given by Lindblom \cite{lindblom:axisym}, assuming that the fluid is
viscous.

Equilibrium states of fluids or collisionless
matter play an important role in astrophysics, providing the basic models of
stars and galaxies. Depending on the equation of state, or in the case of
collisionless matter, on the properties of the distribution function, a
steady state may describe a compact body, or a configuration where the matter
density is nowhere vanishing. Typically, the objects of interest are
compact.

In addition to fluids and collisionless matter, elastic bodies are of
considerable interest in astrophysics in view of the fact that there are strong
theoretical reasons for supposing that neutron stars have a solid crust,
modelled by elastic matter, cf. e.g., \cite{Karlovini:2002fc,
Carter:2006mv,Frauendiener:2007yx}.
The solutions which have been constructed in the just mentioned papers
are all spherically symmetric,
although perturbation analyses have been carried out, allowing for
axial perturbations breaking the spherical symmetry \cite{Karlovini:2007ut}.

\subsection{Rotating bodies in elasticity}
Elastic matter
is, as discussed in section \ref{sec:relelast} below,
described by a map $f^A$ from spacetime to a three dimensional
manifold, called material manifold or body, whose points label
the particles making up the elastic continuum, and which is
taken to be a connected, bounded domain in flat
$\Re^3$.

In considering a rotating steady state, it is important to distinguish
between the microscopic and macroscopic degrees of freedom. The microscopic
degrees of freedom of the elastic matter are described by the configuration
$f^A$, while the macroscopic aspects are described by the stress energy
tensor generated by the matter, and the metric of the spacetime containing
the body. For a rotating body in equilibrium, it is the
case that the stress energy tensor, as well as the spacetime metric
are stationary, i.e. invariant under the
flow of a Killing vector $\xi^\mu \partial_\mu = \partial_t$,
called the stationary Killing
vector, while the matter particles, described by the configuration map, are in
motion relative to $\partial_t$.

As mentioned above, the rotating bodies we construct are axially
symmetric. In particular, the spacetime containing the body admits a Killing
field $\eta^\mu \partial_\mu = \partial_\phi$,
called the axial Killing vector, which commutes with
$\xi^\mu$. In addition, there is a constant $\Omega$, the
angular frequency of rotation, such that the matter particles move along the
helical orbits of the Killing vector $\xi^\mu + \Omega \eta^\mu$,
i.e. the configuration $f^A$ is constant along the flow of the helical
field.

It is nevertheless the case, assuming axisymmetry of the configuration, and
that the elastic material
is isotropic and frame indifferent, cf. section \ref{sec:relelast},
that all spacetime tensors naturally
derived from the configuration are both axisymmetric
and stationary. This applies for example to the matter flow
vector induced by the configuration, the stress tensor or, in
fact, to the full stress energy tensor.

We now briefly describe the method used in this paper.
Consider a Cauchy surface $M$ transverse to the stationary
Killing field. The equations for the gravitational field variables
are derived by imposing the condition of stationarity
and restricting the Einstein equations, reduced in harmonic gauge,
to $M$. No axisymmetry condition is imposed at this stage.

The Einstein equations imply, via the Bianchi identity,
a set of equations for the matter variables. These equations are derived by
considering a configuration which
is comoving with respect to the helical flow, and restricting to $M$. Here
the axial
vector field $\eta^\mu$ is assumed to be specified in advance.
There are, a priori, four matter equations for the three unknowns
$f^A$. We deal with this problem by simply dropping one of the four
equations. It turns out, however, that this
supplementary equation follows from the others when $\eta^\mu$
is Killing, as
is the case for a solution to the system derived by the above procedure.

The resulting coupled system of equations, assuming standard
constitutive conditions for the elastic material, is elliptic
for suffiently small values of $\Omega$. The system depends on the parameters
$G$ and $\Omega$.
We look for solutions to this system for small, nonzero, values of
$G,\Omega$, near the background solution given by taking the spacetime to
be Minkowski, the configuration to be stressfree and the Newton
constant $G$ and $\Omega$ to be both zero.

The boundary between
the matter region and the vacuum region depends on the unknown
configuration. To deal with this problem we write the
equations in material form in a way analogous to \cite{ABS} and
apply the implicit function theorem. This can not be
 done directly due to the
failure of the linearized operator to be invertible. This is a
standard problem for elasticity with natural boundary
conditions, i.e. vanishing normal stress at the boundary. Following
\cite{ABS}, what we actually solve is a
projected version of the field equations, such that the implicit function
theorem does apply. One must then show, as in fact turns out to be the case,
that the solution to this projected system is actually a solution to the full
system.

So far the vector field $\eta^\mu$
was essentially arbitrary except for the condition that it
commute with $\xi^\mu$. In order to ensure that $\eta^\mu$
is a Killing vector, we now assume that the material manifold is
axisymmetric as a subset of Euclidean $\mathbb{R}^3$ and that
$\eta^\mu$ is the pull back of the axial vector field on the
body under the trivial configuration. It then follows by
uniqueness that the vector field $\eta^\mu$ is a Killing
vector.

It now remains to show that the solution found by the
implicit function argument satisfies the Einstein equations. In particular,
the gauge conditions must be satisfied
and the elastic equation must be valid in its original form.
This condition is equivalent to the condition that
a certain linear system of equations
coming from the Bianchi identities has only the trivial
solution. In fact, the linear system under discussion
is homogenous precisely because
the Killing nature of $\eta^\mu$ guarantees that the
above mentioned supplementary equation is satisfied, provided that the
main elastic equation is valid. The rest
of the argument follows essentially the pattern of \cite{ABS}.

\subsection{Outline of the paper}
In section \ref{sec:relelast},
we give some background on relativistic
elasticity.
Section \ref{sec:stationary-metric} introduces stationary metrics and defines
the field variables $h_{ik}, U, \psi_i$ used to parametrize the spacetime
metric.
In contrast to the static case, there is a further
component $\psi_i$ of the gravitational field, corresponding to the failure of
$\xi^\mu$ to be hypersurface orthogonal in general.
Next, in section \ref{sec:reduced},
the stationary
Einstein equations are written in terms of the field variables just introduced, this
corresponds effectively to performing a Kaluza-Klein reduction.

The field equations imply a set of integrability conditions, which are
derived in section \ref{sec:integrability}.
One of these identities
is the elasticity equation,
which is later used as one of the set of equations to be solved
using the implicit function theorem.
The rotation of the elastic body is introduced in section \ref{sec:rotation}.
This is done by choosing a spacelike vector field
$\eta^\mu\partial_\mu$ which commutes with the stationary
Killing vector $\xi^\mu$ and assuming that
$f^A,_\mu (\xi^\mu + \Omega \eta^\mu)$
be zero.
It will later turn out that $\eta^\mu$ is actually the
axial Killing vector.

In section \ref{sec:stress},
the stress tensor is expressed in terms of the geometric
variables and the configuration $f^A$. In particular, this allows us to write
the components of the stress energy tensor in terms of the field
variables, and obtain (\ref{eq:redein1234}),
the basic PDE system for $h_{ik}, U,
\psi_i, f^A$.
As some of the equations are not elliptic we use, as in the
static case, harmonic coordinates to extract an elliptic system.
In the stationary case, it is necessary to make explicit
use also of the condition
that the time function be harmonic, cf. section \ref{sec:gauge}.
Finally, we have reduced the field equations to an elliptic free boundary value
problem in space. In order to avoid dealing directly with the free boundary
aspect of the problem, we move all equations to the body.
This is done in section
\ref{sec:material}, following closely the procedure in \cite{ABS}.
The final details needed to completely specify the PDE problem to be solved
are introduced in section \ref{sec:constitutive}.
There we introduce the relaxed state and a flat metric on the
body. We assume that the shape of the body is axi--symmetric
and use the relaxed configuration to define the vector field
$\eta^i$ in space. We also introduce the assumption that the elastic material
is isotropic.

As in the static case considered in \cite{ABS}, we must consider a projected
system in order to be able to apply the implicit function theorem. The
analytical aspects of this problem are considered in section
\ref{sec:analytical}.
The solution to the projected system is then shown in section
\ref{sec:equilibration} to be a solution to the full set of field equations
in the body frame,
and to be axisymmetric.
In section \ref{sec:eulerian},
we move the projected equations and solutions back to
space. The vector field $\eta^i$ is proved to be a Killing field in section
\ref{sec:equivariance}.

In section \ref{sec:divergence}, we derive some divergence identities play an
essential role in the equilibration argument. This leads up to the main
theorem \ref{thm:mainthm}, which is stated and proved in section
\ref{sec:mainthm}. In particular, we prove that
the harmonic coordinate conditions are satisfied for the solution of the
reduced system that has been constructed, and hence that we have solved the
full set of field equations.

The spacetimes constructed in Theorem \ref{thm:mainthm} have an isometry
group $\Re \times S^1$ generated by the commuting Killing fields $\xi^\mu,
\eta^\mu$. For spacetimes with two commuting Killing fields, it was first
proved by Papapetrou
\cite{papapetrou} in vacuum and by Kundt and Tr\"umper \cite{kundt:truemper}
for fluids, that orthogonal transitivity holds. Recall that orthogonal
transitivity is the condition that the
distribution of 2-surface elements perpendicular to the generators of the
symmetry group is surface forming. This condition
is used for constructing Weyl-type coordinates which play a dominant role in attempts in the exact solution literature
to find rotating body solutions, see \cite{stephani:etal:exact} and
references therein.

In section
\ref{sec:orth-trans},
we establish that for the spacetimes constructed in theorem
\ref{thm:mainthm}, the distribution defined by the 2--surface
elements orthogonal to the group orbit for the action of the
stationary and axisymmetric Killing vector is integrable.
In the case of a smooth spacetime, where the Frobenius theorem applies
directly, this fact implies that orthogonal transitivity holds,
i.e., that there are 2-surfaces perpendicular to the generators
of the symmetry group. In the present case, however, this step needs further
analysis which we defer to a later paper.

\section{The field equations of a rotating, self-gravitating elastic body}
\label{sec:fieldeq}
\subsection{Relativistic elasticity} \label{sec:relelast}
Let $(\MM,g_{\mu\nu})$ be a 3+1 dimensional
spacetime. The {\em body} $\Bo$ is a 3-manifold, possibly with boundary. We shall
consider the case when $\Bo$ is a bounded domain in the extended body
$\Re^3_{\Bo}$. The body domain $\Bo$ is assumed to have a
smooth boundary. In this paper, we shall only consider the case where $\Bo$
is connected. The fields
considered in elasticity are configurations $f : \MM \to \Bo$ and
deformations $\phi: \Bo \to \MM$, with the property that $f \circ \phi =
\id_{\Bo}$.

Let $t$ be a time function on $M$ and introduce a 3+1 split
$\MM = \Re \times M$.
We consider coordinates $(x^\mu) = (t,x^i)$ on
$\MM$, where $x^i$ are coordinates on the space manifold $M$.
On $\Re^3_{\Bo}$
we use coordinates $X^A$. The body
$\Bo$ is
endowed with metric $\delta_{\Bo}$ and a compatible volume form
$V_{ABC}$. We assume that in a suitable Cartesian coordinate system
$\delta_{\Bo}$ has components $\delta_{AB}$ where $\delta_{AB}$ is the
Kronecker delta, and $V_{123} = 1$.

The configuration $f: \MM \to \Bo$ is by assumption
a submersion. The derivative of $f$ is assumed to have a timelike kernel,
i.e. there is a unit timelike vector field $u^\mu$ on $f^{-1}(\Bo)$ with
$u_\mu u^\mu = -1$, such that
$$
u^\mu f^A{}_{,\mu} = 0 \,.
$$
The field $u^\mu$ is the velocity field of the matter and describes the
trajectories of the material particles.

Let $\Lambda = \Lambda(f, \partial f , g)
$ be the energy density for the elastic material in its own rest
frame.
The Lagrange density
for the self-gravitating elastic body now takes the
form
$$
\LL =  - \frac{R_g \sqrt{-g}}{16\pi G} + \Lambda \sqrt{-g} \,.
$$
The Einstein equations resulting from the variation of the action with
respect
to $g^{\mu\nu}$ take the form
$$
G_{\mu\nu} = 8 \pi G T_{\mu\nu} \,,
$$
where $G_{\mu\nu}$ is the Einstein tensor
of $g_{\mu\nu}$ and $T_{\mu\nu}$ is the
stress energy tensor of the material, given by
$$
T_{\mu\nu} = 2\frac{\partial \Lambda}{\partial g^{\mu\nu}} - \Lambda
  g_{\mu\nu} \,.
$$
On the other hand, the canonical stress energy tensor is given by
$$
\TT_\mu{}^\nu = \frac{\partial \Lambda}{\partial f^A{}_{,\nu}} f^A{}_{,\mu} -
\delta_\mu{}^\nu \Lambda \,.
$$
General covariance implies, by the Rosenfeld-Belinfante theorem, that
$$
\TT_{\mu\nu} = - T_{\mu\nu} \,,
$$
see \cite[section 7]{KM}.

Given a configuration $f^A(x^\mu)$,
define $\gamma^{AB} = f^A{}_{,\mu} f^B{}_{,\nu} g^{\mu\nu}$ and let
$\gamma_{AB}$ be the inverse of $\gamma^{AB}$. General
covariance implies
$\Lambda$ is of the form $\Lambda = \Lambda(f^A, \gamma^{AB})$, cf.
\cite[section 7]{KM}, see also \cite[section 4]{BS:CQG2003}.
A stored energy function of this form is said to satisfy material frame
indifference.
If in addition, as we shall later assume, $\Lambda$
depends only on the principal invariants $\lambda_i$, $i=1,2,3$ of
$\gamma^{AB}$, defined as the elementary symmetric polynomials in the
eigenvalues of $\gamma^A{}_B = \gamma^{AC} (\delta_{\Bo})_{CB}$, then the
material is called isotropic.

Define
$$
S_{AB} = 2\frac{\partial \Lambda}{\partial \gamma^{AB}} -  \Lambda
\gamma_{AB} \,.
$$
Then we have
\begin{equation}\label{eq:TmunuLam}
T_{\mu\nu} = \Lambda u_\mu u_\nu + S_{\mu\nu} \,,
\end{equation}
where $S_{\mu\nu} = S_{AB} f^A{}_{,\mu} f^B{}_{,\nu}$. In particular
$S_{\mu\nu} u^\nu = 0$.
The relativistic
number density $n_g$ is defined by
$$
 n_g^2 = \frac{1}{3!} V_{ABC} V_{A'B'C'} \gamma^{AA'}\gamma^{BB'}
\gamma^{CC'} \,.
$$
We have $n_g = (\det \gamma^{AB})^{1/2}$ and hence
\begin{equation}\label{eq:dndgam}
\frac{\partial n_g}{\partial \gamma^{AB}} = \half n_g \gamma_{AB} \,.
\end{equation}
Define the stored energy function $\epsilon$ by
\begin{equation}\label{eq:Lam-ng-eps}
\Lambda = n_g \epsilon \,,
\end{equation}
and the elastic stress tensor $\tau_{AB}$ by
$$
\tau_{AB} = 2 \frac{\partial \epsilon}{\partial \gamma^{AB}} \,.
$$
With these definitions, $S_{AB}$ takes the form
$$
S_{AB} = n_g \tau_{AB} \,,
$$
and we can write
$$
T_{\mu\nu} = n_g \epsilon u_\mu u_\nu + n_g \tau_{AB} f^A{}_{,\mu} f^B{}_{,\nu}
\,.
$$
See \cite{shadi:elast} for a more explicit expression of $T_{\mu\nu}$ in
terms of the invariants $(\lambda_i)$.

If material frame indifference holds, then if $\Lambda$ is
viewed as a
functional of $f^A, g_{\mu\nu}$, we have that for any spacetime
diffeomorphism $\sigma$,
$$
\Lambda[f \circ \sigma, \sigma^* g] = \Lambda[f,g]\circ \sigma \,,
$$
and hence all spacetime quantities constructed from
$\Lambda,f^A,g_{\mu\nu}$ are covariant under $\sigma$, including
$n_g$, $u^\mu$ and $\tau_{AB} f^A{}_{,\mu} f^B{}_{,\nu}$. In particular, this
holds also for $T_{\mu\nu}$.

Let $\Sigma$ be an isometry of $(\Bo,\delta_{\Bo})$.
The matrix $(\Sigma_* \gamma)^A{}_B$ is related to $\gamma^A{}_B$ by an
orthogonal similarity transform and hence has the same
invariants $\lambda_i$ as $\gamma^A{}_B$.
Hence, for an isotropic material,
$$
\Lambda[\Sigma \circ f, g] = \Lambda[f, g] .
$$

\subsection{Material and spacetime isometries}
\label{sec:material:isometry}
We now introduce the notion of symmetry of a configuration which will play an
important role in this paper.
Suppose the spacetime $(\MM,g)$ has an isometry $\sigma$. Then $\sigma$ defines
a material symmetry of  $f^A$ if there is an isometry
$\Sigma$ of $(\Bo, \delta_{\Bo})$ such that
$$
\Sigma \circ f = f \circ \sigma \,.
$$
Thus, in particular, if the configuration is comoving with an
isometry, i.e., if $u^\mu$ is proportional to a Killing vector
$\xi^\mu$, then the configuration has the flow $\sigma_s$ of
$\xi^\mu$ as a material isometry. with $\Sigma$ given by the
identity map on $\Bo$, in which case it follows that $f^A \circ
\sigma_s = f^A$. However, this does not hold for a general
one-parameter family of material isometries. It follows from the
last two statements in the previous subsection 
that, for an isotropic material,
a spacetime
isometry $\sigma$ which also defines
a material isometry leaves the Lagrangian invariant, i.e. $\Lambda
[f,g] = \Lambda [f,g] \circ \sigma$, and thus $T_{\mu \nu}$ is also
invariant under $\sigma$, i.e. $\sigma^* T = T$.

The following is an
example which is relevant for the situation in this paper.
Suppose we have two timelike Killing vectors $\xi^\mu$ and $\xi'^\mu$. In the
situation considered in this paper, the interesting case is where $\xi^\mu$
is the stationary Killing field, while ${\xi'}^\mu = \xi^\mu + \Omega
\eta^\mu$ is the helical Killing field. Then one may consider the case where
the configuration is comoving with respect to ${\xi'}^\mu$
while the flow
$\sigma_s$ of $\xi^\mu$ defines a configuration symmetry in the sense that
there is a one-parameter family of isometries $\Sigma_s$ of the body
such that
$$
\Sigma_s \circ f^A = f^A \circ \sigma_s \,.
$$
In this case, $\Sigma_s$ are rotations of the body.
We see from the above that it is possible for the
configuration to explicitly depend on the Killing time $t$, defined with
respect to $\xi^\mu$, although $T_{\mu\nu}$ is independent of $t$.

\subsection{Stationary metrics}\label{sec:stationary-metric}
We now assume $(\MM, g)$ is stationary, i.e. there is a timelike
Killing field $\xi^\mu \partial_\mu = \partial_t$. Further we assume the
space manifold $M$ is diffeomorphic to $\Re^3$. It will sometimes be
convenient to denote this space by $\Re^3_{\Sp}$.
\mymnote{LA: discuss here properties of $f^A{}_{,i}$ and $\phi^i{}_{,A}$, for
  example $f^A{}_{,i} \phi^i{}_{,B} = \delta^A{}_B$ etc.}
Define
a function $U = \half \ln \xi^\mu \xi_\mu$ and a one-form $\psi = \psi_i
dx^i$ such that $e^{-2U} \xi_\mu dx^\mu = dt +  \psi$. Then $g$ can
be written in the form
\begin{equation}\label{metric}
g_{\mu\nu}\,dx^\mu dx^\nu = - e^{2 U}(dt +  \psi_i dx^i)^2 +
e^{-2U}h_{ij}dx^i dx^j \,,
\end{equation}
where $h_{ij} dx^i dx^j$ is a metric on the level sets of $t$, and $U,
\psi_i, h_{ij}$ are time independent. The inverse
metric takes the form
\begin{equation}\label{imetric}
g^{\mu\nu} \partial_\mu \partial_\nu = - e^{-2U} \partial_t^2 +
e^{2U}h^{ij}(\partial_i - \psi_i
\partial_t)(\partial_j -  \psi_j \partial_t)\;,
\end{equation}
where $h^{ij}h_{jk} = \delta^i{}_k$. The spacetime volume element is given by
\begin{equation}\label{det}
\sqrt{-g}=e^{-2U}\sqrt{h} \,.
\end{equation}
The assumption that $\xi^\mu \partial_\mu = \partial_t$ is a Killing vector
implies
\begin{equation}\label{harm}
\Box_g t = - e^{2U} D^i \psi_i \quad
\Box_g x^i = - e^{2U} h^{jk}\Gamma^i_{jk}\;,
\end{equation}
where $D_i$ and the Christoffel symbols refer to $h_{ij}$.

\subsection{Kaluza-Klein reduction} \label{sec:reduced}
Let $\omega_{ij} = \partial_{[i}\psi_{j]}$. The scalar curvature $R_g$ for a
metric of the form (\ref{metric}) is given by
\begin{equation} \label{eq:scal}
R_g \sqrt{-g} = \sqrt{h} \left (R_h + 2\Delta_h U - 2 |D U|_h^2
+  e^{4 U}  |\omega|_h^2 \right ) .
\end{equation}
Here $|DU|_h^2 = D_k U D^k U$ and similarly for $|\omega|_h^2$.
Define $H^{AB}$ by
$$
\gamma^{AB} = e^{2U} H^{AB} \,.
$$
The reduced number density $n$ is defined with respect to $H^{AB}$,
$$
n^2 = \frac{1}{3!}
V_{ABC} V_{A'B'C'} H^{AA'}H^{BB'}H^{CC'} \,.
$$
Then we have
\begin{equation} \label{eq:dndgamred}
\frac{\partial n}{\partial H^{AB}} = \half n H_{AB} \,,
\end{equation}
\begin{equation} \label{eq:ngnform}
n_g = e^{3U} n \,,
\end{equation}
so that with the form (\ref{metric}) for $g$,
we have
\begin{equation} \label{eq:Lam-n-eps}
\Lambda \sqrt{-g}
= n \epsilon e^U \sqrt{h} \,.
\end{equation}
Taking into account the fact that
the term $2\Delta_h U$ in the scalar curvature
expression (\ref{eq:scal}) contributes a total divergence to the action and can
be dropped, we may now write the action in the reduced form
\begin{align*}
\LL = - \frac{\sqrt{h}}{16\pi G} \left (R_h  - 2 |D U|_h^2
+ e^{4 U}  |\omega|_h^2 \right ) + \rho \, e^U \sqrt{h} \,,
\end{align*}
where $\rho = n\epsilon$.
Let $G_{ij} = R_{ij} - \half
R h_{ij} $ be the Einstein tensor of $h_{ij}$ and define
\begin{equation}\label{eq:Theta}
\Theta_{ij} = \frac{1}{4 \pi G}[(D_i U) (D_j U) - \frac{1}{2} h_{ij}\, (D_k
  U)(D^k U)] \,,
\end{equation}
and
\begin{equation}\label{eq:Omega}
\Omega_{ij} = \frac{1}{4 \pi G}\,e^{4U}[\,  \frac{1}{4}
h_{ij}\,\omega_{kl}\omega^{kl} - \omega_{ik}\omega_j{}^k] \,.
\end{equation}
The reduced field equations now take the form
\mymnote{LA: note sign in (\ref{eq:Domega})}
\begin{subequations}\label{eq:rhofield}
\begin{align}
\Delta_h U  &= 4\pi G e^U \left ( \rho +
\frac{\partial \rho}{\partial U} \right )\chi_{f^{-1}(\Bo)} - e^{4U}
\omega_{kl}\omega^{kl} \label{eq:DeltaU} \,,
\\
D^i (e^{4U} \omega_{ij} ) &= - 8\pi G e^U \frac{\partial \rho}{\partial \psi^j}
\chi_{f^{-1}(\Bo)} \,, \label{eq:Domega}
   \\
G_{ij} &= 8\pi G\left (
\Theta_{ij} + \Omega_{ij} + e^U (2 \frac{\partial \rho}{\partial
  h^{ij}} - \rho h_{ij} )  \chi_{f^{-1}(\Bo)} \right ) \,. \label{eq:Gij}
\end{align}
\end{subequations}
In (\ref{eq:rhofield}) we have used the indicator function
$\chi_{f^{-1}(\Bo)}$ of the body to make explicit the support of $\rho$.
Define $\tau, \tau_i, \tau_{ij}$ by
\begin{equation}\label{eq:Tmunu-tau}
T_{\mu\nu} = \tau (dt + \psi_i dx^i)^2 + 2 \tau_j dx^j (dt + \psi_i dx^i) +
\tau_{ij} dx^i dx^j \,.
\end{equation}
Then we have
\begin{lemma} \label{lem:rhotau}
\begin{subequations}\label{eq:rhotau}
\begin{align}
e^U  (2 \frac{\partial \rho}{\partial
  h^{ij}} - \rho h_{ij} ) &= \tau_{ij} \,,\\
e^U \frac{\partial \rho}{\partial \psi^i} &= - \tau_i \,,\\
e^U (\rho + \frac{\partial \rho}{\partial U} )
&= e^{-4U} \tau + \tau_\ell{}^\ell \,.
\end{align}
\end{subequations}
\end{lemma}
For proof of Lemma \ref{lem:rhotau}, see appendix \ref{sec:rhotauproof}.
After substituting (\ref{eq:rhotau}) into  (\ref{eq:rhofield}) the reduced
field equations take the form\footnote{Equation (\ref{ein2}) corrects a
  typo in   \cite[(2.47)]{beig:schmidt:timeindep}.}
\begin{subequations}\label{ein123}
\begin{alignat}{1}
\label{ein1}
\Delta_h U & =  4 \pi G \,\chi_{f^{-1}(\Bo)}(e^{-4 U} \tau + \tau{}_k{}^k) -
e^{4U} \omega_{kl}\omega^{kl} \,, \\
\label{ein2}
D^i(e^{4U} \omega_{ij}) & =  8 \pi G \,\chi_{f^{-1}(\Bo)}\tau_j \,, \\
\label{ein3} G_{ij} & =  8 \pi G
(\chi_{f^{-1}(\Bo)}\tau_{ij} + \Theta_{ij} + \Omega_{ij}) \,.
\end{alignat}
\end{subequations}
\subsection{Integrability conditions} \label{sec:integrability}
The quantities $\Theta_{ij}$ and $\Omega_{ij}$ satisfy the identities
\begin{equation}\label{id1}
8 \pi G\, D^j \Theta_{ij} = 2 \,(D_i U)\Delta_h U \,,
\end{equation}
and, using $D_{[i} \omega_{jk]}=0$,
\begin{equation}\label{id2}
8 \pi G\, D^j \Omega_{ij} = 2 \,[e^{4U} (D_i U)
\omega_{kl}\omega^{kl} - \omega_{ik}\,D^j(e^{4U}\omega_j{}^k)] \,.
\end{equation}
In case $G \ne 0$, we obtain from (\ref{ein2}) the integrability conditions
\begin{equation}\label{int1}
D^i \tau_i = 0 \,,
\end{equation}
together with the boundary condition
\begin{equation}\label{bound1}
\tau_i n^i|_{f^{-1}(\partial \Bo)}=0 \,.
\end{equation}
Further, we have
\begin{equation}\label{int2}
D^j \tau_{ij} - 2\, \omega_{ij} \tau^j = - (D_i U) (e^{-4 U} \tau +
\tau_k{}^k) \,,
\end{equation}
and
\begin{equation}\label{bound2}
\tau_{ij} n^j|_{f^{-1}(\partial \Bo)}=0 \,,
\end{equation}
as a consequence of
the contracted Bianchi identities for $h_{ij}$ applied to the left hand side
of (\ref{ein3}), together with (\ref{id1}), (\ref{id2})
and (\ref{ein1}).

\mymnote{LA: Why is the remark on the number of boundary conditions
  commented out??}

\subsection{Implementing rotation} \label{sec:rotation}
Define a vector field $\eta^\mu$ by
\begin{equation}\label{eta}
\eta^\mu \partial_\mu = \eta^i \partial_i \,.
\end{equation}
The scalar product $\alpha = g_{\mu\nu}\xi^\mu \eta^\nu$
satisfies
\begin{equation}\label{scalar}
e^{-2U}\alpha + \psi_j \eta^j = 0 \,.
\end{equation}
Since by assumption $(\MM, g)$ is stationary with respect to $\xi^\mu$, it
holds that $\eta^\mu$
commutes with $\xi^\mu$ if and only if $\eta^i$ does not depend on $t$.
In particular, in the case we are considering, the
vector field $\eta^\mu$ is itself a
Killing vector if and only if the equations
\begin{subequations}\label{kilUpsih}
\begin{alignat}{1}
\label{kilU} \Lie_\eta U & = 0 \,,\\
\label{kilpsi} \Lie_\eta \psi_i & = 0 \,,\\
\label{kilh} \Lie_\eta h_{ij} & = 0 \,,
\end{alignat}
\end{subequations}
hold. In these expressions the operator $\Lie_\eta$ means the
Lie derivative of the respective object with respect to $\eta^k \partial_k$.
Note that (\ref{kilpsi}) implies
\begin{equation}\label{kilpsi2}
2 \,\omega_{ij} \eta^j + D_i(e^{-2U} \alpha) = 0 \,.
\end{equation}
Define the velocity field $u^\mu$ by
\begin{equation}\label{eq:u-def}
u^\mu = b^{-1} (\xi^\mu + \Omega \eta^\mu) \,,
\end{equation}
where $\Omega$ is a real parameter corresponding to the rotation speed, and
$b$ is a normalizing factor, determined by $u^\mu u_\mu = -1$ in
$f^{-1}(\Bo)$.
It is important to note here that the rotational field
$\xi^\mu + \Omega \eta^\mu$ in general will fail to be globally timelike for
nonzero values of $\Omega$. However, for a suitable range of $\Omega$, it makes
sense to require $\xi^\mu + \Omega \eta^\mu$ to be timelike in the body.

We now impose rotation of the body by requiring that
the configuration $f^A$ satisfies the condition
$u^\mu f^A{}_{,\mu} = 0$, i.e.
\begin{equation}\label{eq:hel}
f^A{}_{,\mu} (\xi^\mu + \Omega \,\eta^\mu) = 0 \,.
\end{equation}
Since $T_{\mu\nu} u^\nu = - \rho u_\mu$, due to (\ref{eq:TmunuLam})
the stress energy tensor satisfies the relation
\begin{equation}\label{elast}
u_{[\mu} T_{\nu]\rho} u^\rho = 0 \,.
\end{equation}
It follows from (\ref{elast}), that
\begin{equation}\label{eigen}
(u_0 T_{i\mu} - u_i T_{0\mu})u^\mu = 0
\end{equation}
holds,
which, using (\ref{eq:hel}), after some cancellations and multiplying by
$e^{-2U}$  gives
\begin{multline}\label{master}
(1 - \Omega e^{-2U} \alpha)^2 \tau_i + \Omega (1 - \Omega e^{-2
U}\alpha)\tau_{ij}\eta^j \\
+ \Omega e^{-4U} \eta_i[(1 - \Omega e^{-2
U} \alpha)\tau + \Omega \tau_j \eta^j] = 0 \,.
\end{multline}
Equation (\ref{master}) can be proved by explicit computation, using (\ref{eq:rhotau}) and (\ref{eq:gammaform}).
As a consequence of (\ref{master}) we have
\begin{lemma}\label{lem:master1} For sufficiently small $\Omega$,
\begin{equation}\label{master1}
(1 - \Omega e^{-2U} \alpha) \tau_i + \Omega \tau_{ij}\eta^j + \Omega
e^{-4U} \eta_i[\tau + \Omega (1 - \Omega e^{-2U} \alpha)^{-1} \tau_j
\eta^j] = 0 \,.
\end{equation}
\end{lemma}

\subsection{Stress tensor} \label{sec:stress}
In order to write the field equations, we shall need the stress tensor for
the elastic material. For consistency with the treatment of the static case
considered in \cite{ABS}, we shall here make use of an analogous form of the
stress tensor.
Recall that assuming material frame indifference,
the stored energy function $\epsilon$ is a function of
$f^A$ and $\gamma^{AB} = f^A{}_{,\mu} f^B{}_{,\nu} g^{\mu\nu}$.
Taking equation (\ref{eq:hel}) into account, we find
\begin{equation}\label{eq:gammaform}
 \gamma^{AB} = f^{(A}{}_{,i}f^{B)}{}_{,j}[e^{2U} h^{ij} + 2 \Omega e^{2U} \psi^i \eta^j + \Omega^2
 (e^{2U} \psi^k \psi_k - e^{-2U})\eta^i \eta^j]\;.
\end{equation}
In the computations below we shall make use of $H^{AB}$ defined
by $\gamma^{AB} = e^{2U} H^{AB}$, so that
$$
H^{AB} =  f^{(A}{}_{,i}f^{B)}{}_{,j}[ h^{ij} + 2 \Omega  \psi^i \eta^j + \Omega^2
 ( \psi^k \psi_k - e^{-4U})\eta^i \eta^j] \,.
$$

Let
\begin{equation}\label{eq:stress}
\sigma_{AB}=-2 \frac{\partial \epsilon}{\partial H^{AB}} \,, \quad
\sigma_{\mu\nu}= n f^A{}_{,\mu}f^B{}_{,\nu} \sigma_{AB}\,, \quad
\sigma_\mu {}^A=f^B{}_{,\mu}\sigma_{BC} H^{CA} \,.
\end{equation}
It follows from the definition that
$$
\sigma_{AB} = -2 e^{-2U} \frac{\partial \epsilon}{\partial \gamma^{AB}} \,.
$$
Our next task is to evaluate the dependence on $\Omega$ of the terms occurring
in the left hand side of (\ref{eq:rhotau}).
It is straightforward to verify that the following Lemma holds.
\begin{lemma} \label{lem:stressOmega}
There are $z, z_i, z_{ij}$ depending smoothly an $f^A$, $g_{\mu\nu}$, and
their first derivatives, as well as $\Omega$ and $G$,
such that the following equations are valid.
\begin{subequations} \label{eq:stressOmega}
\begin{align}
e^U (2 \frac{\partial \rho}{\partial
  h^{ij}} - \rho h_{ij} ) &= - e^U (\sigma_{ij} - \Omega z_{ij} ) \,,\\
e^U \frac{\partial \rho}{\partial \psi^i} &= - e^U \Omega z_i \,, \\
e^U (\rho + \frac{\partial \rho}{\partial U} )
&= e^U( n \epsilon - \sigma_\ell{}^\ell) + \Omega z) \,.
\end{align}
\end{subequations}
\end{lemma}
By the results of Lemma \ref{lem:stressOmega} and Lemma \ref{lem:rhotau},
we have
$$
\tau_{ij} = - e^U(\sigma_{ij} - \Omega z_{ij}) \,.
$$
We are now able to rewrite the integrability condition (\ref{int2}) in the
form
$$
D^j (e^U \sigma_{ij}) = e^U(n \epsilon - \sigma_\ell{}^\ell) D_i U + \Omega
[ D^j (e^U z_{ij}) + 2  e^U \omega_{ij}  z^j +  z D_i U] \,.
$$
Taking the above facts into account, we arrive at the  system of
equations
\begin{subequations}\label{eq:redein1234}
\begin{align}
\Delta_h U &=  4 \pi G \,\chi_{f^{-1}(\Bo)}
e^U( n \epsilon - \sigma_\ell{}^\ell) + \Omega z)
- e^{4U} \omega_{kl}\omega^{kl} \,,
\label{eq:redein1}
\\
D^i(e^{4U} \omega_{ij}) & =  8 \pi G \,\chi_{f^{-1}(\Bo)}e^U \Omega z_j \,,
\label{eq:redein2}
\\
G_{ij} &=  8 \pi G
[- \chi_{f^{-1}(\Bo)}e^U(\sigma_{ij} - \Omega z_{ij})
+ \Theta_{ij} + \Omega_{ij}] \,,
\label{eq:redein3}
\\
D^j (e^U \sigma_{ij}) &= e^U(n \epsilon - \sigma_\ell{}^\ell) D_i U \\
&\quad + \Omega
[ D^j (e^U z_{ij}) + 2  e^U \omega_{ij}  z^j +  z D_i U], \quad \text{ in }
f^{-1}(\Bo) \,, \label{eq:redein4} \\
\intertext{subject to the boundary condition}
&\quad (\sigma_{ij} - \Omega z_{ij}) n^j \big{|}_{\partial f^{-1}(\Bo)} = 0 \,.
\label{eq:redein5}
\end{align}
\end{subequations}
\subsection{Gauge reduction} \label{sec:gauge}
Two of the equations in the system (\ref{eq:redein1234}) fail to be elliptic
in the form given above, namely (\ref{eq:redein2}) and (\ref{eq:redein3}). The
reason for this failure is the related to the diffeomorphism invariance of
the 4-dimensional Einstein equations. As in the static case, the method which
shall be used to avoid this problem is make use of harmonic coordinates.

Let $\square$ denote the wave operator in $(\MM, g)$. Taking into account of
the fact that $g_{\mu\nu}$ is stationary, we have
\begin{align}
\square t &= \frac{1}{\sqrt{-g}} \partial_\mu (g^{\mu\nu} \sqrt{-g}
\partial_\nu ) t \\
&= e^{2U} D^i \psi_i \,.
\end{align}
Thus, $e^{2U} D^i \psi_i = 0$ precisely when the time $t$ is harmonic.

The left hand side of equation (\ref{eq:redein2}) is of the form
\begin{equation}\label{eq:psi-ellip}
D^i (e^{4U} \omega_{ij}) =  e^{4U} [ 4 D^i U \omega_{ij} +
\half  ( \Delta \psi_j - R_j{}^k \psi_k) ] - \half e^{4U} D_j D^i \psi_i \,.
\end{equation}
The term $D_j D^i \psi_i$ causes this expression to fail to be an elliptic
in $\psi_i$.
However, the following reduced form of equation (\ref{eq:redein2}),
\begin{equation}\label{eq:reduced-psi}
D^i(e^{4U} \omega_{ij}) + \half D_j (e^{4U} D_i \psi^i) =  8 \pi G
\,\chi_{f^{-1}(\Bo)}e^U \Omega z_j \,,
\end{equation}
which modifies the left hand side by a quantity that vanishes if the harmonic
time condition is satisfied, is elliptic in $\psi_i$.

Similarly, (\ref{eq:redein3}) fails to be elliptic due to
the covariance of $R_{ij}$.
Following \cite[section 3.1]{ABS}, let $V^i = h^{jk} (\Gamma^i_{jk} - \widehat
\Gamma^i_{jk})$ where $\widehat \Gamma^i_{jk}$ are the Christoffel symbols of
a fixed Euclidean background metric on $M$. Then $V^i = 0$ is the condition
for harmonic coordinates in $M$.
By replacing $R_{ij}$ by $R_{ij} - D_{(i} V_{j)}$ we arrive, after rewriting
equation (\ref{eq:redein3}) making use of the identity \cite[(3.11)]{ABS}
at the reduced Einstein equation
\begin{multline}\label{eq:reduced-h}
-\frac{1}{2}\Delta_h h_{ij} + Q_{ij}(h,\partial h) =
- 8 \pi
G e^U (\sigma_{ij}  - h_{ij}\,\sigma_l{}^l
+ \Omega (z_{ij} - h_{ij}
z_l{}^l))\chi_{f^{-1}(\Bo)} \\
+ 2 D_i U D_j U
+ e^{4U} [ h_{ij} \omega_{kl} \omega^{kl} - 2 \omega_{ik} \omega_j{}^k ] \,.
\end{multline}
As in \cite{ABS}, we shall first solve the reduced system involving
(\ref{eq:reduced-h}) and (\ref{eq:reduced-psi}) and once the solution is in
hand show that the solution to the reduced system is actually a solution to
the full system. We construct solutions by an implicit function theorem
argument applied to a projected version of the field equations in material
form.

\subsection{Field equations in material form} \label{sec:material}
In the Eulerian picture, the domain $f^{-1}(\Bo)$ depends on the
unknown configuration $f$. This introduces a ``free boundary'' aspect in the
Eulerian version of the field equations, which we will avoid by passing to
the material, or Lagrangian form of the field equations. In this form of the
equations, the configuration $f$ is replaced by the deformation $\phi$, and
the entire system of field equations is moved to the extended body
$\Re^3_{\Bo}$. In particular, in this formulation, the elastic field equation
lives on the fixed domain $\Bo$.

The Piola transform of $\sigma_i{}^j$ is
$$
\bar \sigma_i{}^A = J ( f^A{}_{,j} \sigma_i{}^j ) \circ \phi \,.
$$
Similarly, we introduce the Piola transform of $z_{ij}$.
Since $\Bo$ has a smooth boundary, there is a linear extension operator which
takes functions on $\Bo$ to functions on $\Re^3_\Bo$. In particular this
allows us to define an extension $\hphi$ of $\phi$ which is equal to $\ii$
outside a compact set. We use $\hphi$ to move the fields from space to
$\Re^3_\Bo$, and use the bar notation introduced in \cite[section 3.2]{ABS}
to denote the quantities transported under $\hphi$.
In particular, we define
$$
\overline{U} = U \circ \hphi \,,
\quad \overline{\partial_i U} = \partial_i U \circ \hphi\,,
\quad \overline{\psi_i} = \psi_i \circ \hphi\,,
\quad \overline{h_{ij}} = h_{ij} \circ \hphi \,.
$$
Note that for the barred quantities, it is the frame components which are
pulled back, and not the tensor itself.
Equation (\ref{eq:redein1}) in the material frame becomes
\begin{equation}\label{eq:redein1-mat}
\overline{\Delta_h U} =  4 \pi G \chi_{\Bo}
e^{\bU}( n \bar \epsilon - \bar \sigma_\ell{}^\ell) + \Omega \bar z)
- e^{4\bU} \overline{\omega_{kl}\omega^{kl}} \,, \quad \text{ in }
\Re^3_{\Bo} \,.
\end{equation}
We remark that covariance of the Laplacian gives
$$
\overline{\Delta_h U} = \Delta_{\hphi^* h} (U \circ \hphi) \,.
$$
Next, equation (\ref{eq:reduced-psi}) in the material frame becomes
\begin{equation}\label{eq:reduced-psi-mat}
\overline{D^i (e^{4U} \omega_{ij})}
 + \half \overline{D_j (e^{4U} D_i \psi^i)}
=
8 \pi G
\,\chi_{\Bo}e^{\bU} \Omega \bar z_j \,.
\end{equation}
Equation (\ref{eq:reduced-h}) becomes
\begin{multline}\label{eq:reduced-h-mat}
-\frac{1}{2}\overline{\Delta_h h_{ij}} + Q_{ij}(\bar h,\partial \bar h) =
- 8 \pi
G e^{\bU} (\bar \sigma_{ij}  - \bar h_{ij}\,\bar \sigma_l{}^l
+ \Omega (\bar z_{ij} - \bar h_{ij}
\bar z_l{}^l))\chi_{\Bo} \\
+ 2 \overline{D_i U D_j U }
+ e^{4\bU} [ \bar h_{ij} \overline{\omega_{kl} \omega^{kl}} - 2
-\overline{\omega_{ik} \omega_j{}^k} ] \,.
\end{multline}
Equations (\ref{eq:redein4}) and (\ref{eq:redein5}) become in the material
frame
\begin{subequations} \label{eq:redein45-mat}
\begin{align}
D_A (e^{\bU} \sigma_i{}^A) &= e^{\bU}(\epsilon - \frac{\bar
  \sigma_\ell{}^\ell}{\bar n} ) \overline{\partial_i U} \nonumber \\
&\quad + \Omega
[ D_A (e^{\bU} \bar z_i{}^A) + 2  e^{\bU} \frac{\overline{\omega_{ij}
  z^j}}{\bar n} +
\bar z \overline{\partial_i U}] \,, \quad \text{ in }
\Bo \,, \label{eq:redein4-mat} \\
\intertext{subject to the boundary condition}
&\quad (\bar \sigma_i{}^A - \Omega \bar z_i{}^A) n_A \big{|}_{\partial \Bo} =
  0 \,.
\label{eq:redein5-mat}
\end{align}
\end{subequations}

\subsection{Constitutive conditions} \label{sec:constitutive}
Similarly to the static case, we shall assume the existence of a relaxed
reference configuration for the elastic material, which is such that suitable
ellipticity properties hold for the elasticity operator evaluated in the
relaxed state.
The relaxed state is given
by the body $\Bo$, a compact, connected
domain $\Bo \subset \Re^3_{\Bo}$ with smooth
boundary $\partial \Bo$, together with a reference configuration $\ii :
\Re^3_\Bo \to \Re^3_\Sp$.
We assume a reference Euclidean metric
$\hat{\delta}$ on $M = \Re^3_{\Sp}$ is given.
The body metric $\Re^3_{\Bo}$ on $\Re^3_{\Bo}$
is defined by
$\delta_{\Bo} = \ii^* \hat{\delta}$. The relaxed nature of the reference
configuration is expressed by the condition
$$
\left(\frac{\partial \epsilon}{\partial H^{AB}} \right) \bigg{|}_{(U =
  0,H=\delta_\Bo)} = 0 \,,
\quad \text{ in }\Bo \,.
$$
The specific rest mass, i.e. the rest mass term in the relativistic stored
energy function, should obey
$$
\mathring{\epsilon} (X) = \epsilon \big{|}_{(U=0,
H = \delta_{\Bo}
)} \geq C \,, \\
$$
for some constant $C > 0$. Further, we assume that the elastic material is
such that there is a constant $C' > 0$ such that the pointwise stability
condition
\begin{equation}\label{constnew2}
\mathring{L}_{ABCD} N^{AB} N^{CD} \geq C' \, (\delta_{CA}\delta_{BD} +
\delta_{CB}\delta_{AD})N^{AB}N^{CD} \,,
\quad \text{ in } \Bo \,,
\end{equation}
holds, where
\begin{equation}\label{constnew3}
\mathring{L}_{ABCD}(X) := \left(\frac{\partial^2 \epsilon}{\partial
  H^{AB}\partial H^{CD}}\right)\bigg{|}_{(U=0,H=\delta_{\Bo})} \,.
\end{equation}
In the isotropic case considered 
in this paper, $\epsilon$ depends only on the invariants of $\gamma^{AB} = e^{2U}
H^{AB}$, defined with
respect to the body metric $(\delta_{\Bo})_{AB}$, cf. section 
\ref{sec:relelast}. It follows that 
$\mathring{\epsilon}$ is 
independent of $X$ and there are
constants $\mathring{\lambda}$ and $\mathring{\nu}$ so that
\begin{equation}\label{lame}
\mathring{L}_{ABCD} = \mathring{\lambda} \delta_{AB}\delta_{CD} + 2
\mathring{\mu}\delta_{C(A}\delta_{B)D}  \,,
\end{equation}
in terms of which the condition (\ref{constnew2}) holds exactly when
\begin{equation}\label{lame1}
\mathring{\mu} > 0 \,, \quad 3 \mathring{\lambda} + 2 \mathring {\mu} > 0  \,,
\end{equation}
cf. \cite[section 4.3]{hughes:marsden}. The constants
$\mathring{\lambda}$ and $\mathring{\mu}$ are apart from a common
constant factor the classical Lam\'{e} moduli. The inequalities
(\ref{lame1}) are usually expressed by saying that the Poisson ratio
defined by $\nu = \frac{\mathring{\lambda}}{2 (\mathring{\lambda} +
\mathring{\mu})}$ satisfy $- 1 < \nu < \frac{1}{2}$. In fact for
most materials occuring in practice there holds $\frac{1}{4} < \nu <
\frac{1}{3}$. \mymnote{LA: add some reference for this, Bobby:
material added, still no citable reference}

We shall assume that the body is axisymmetric. To make this notion concrete,
let $x^i$ and
$X^A$ be coordinates on
$\Re^3_{\Sp}$ and $\Re^3_{\Bo}$, respectively,
so that
the Euclidean metrics
$\hat{\delta}$ and
$\delta_{\Bo}$
have components $\delta_{ij}$ and
$\delta_{AB}$, respectively.
The body
$\Bo$ is axially symmetric if there is a one-parameter subgroup of
Euclidean motions, defined with respect to $\delta_{AB}$, which leaves $\Bo$
invariant. We may without loss of generality assume that the
subgroup leaving $\Bo$
invariant is generated by the Killing field
\begin{equation}\label{eq:eta-Bo-def}
\eta^A \partial_A = X^2 \partial_1 - X^1 \partial_2 \,,
\end{equation}
which necessarily is such that $\eta^A$ is tangent to $\partial \Bo$.
Given the axial Killing field $\eta^A$ on $\Re^3_{\Bo}$,
define a vector field $\eta^i$ on $\Re^3_{\Sp}$ by
\begin{equation}\label{eq:axi}
\eta^i \partial_i = \ii_* (\eta^A \partial_A) \,.
\end{equation}
In particular, we may without loss of generality assume $\eta^i \partial_i$
to be of the form $\eta^i \partial_i = x^2 \partial_1 - x^1 \partial_2$.

In addition to the above mentioned conditions, we shall in the following
assume
that the elastic material is isotropic, cf. section \ref{sec:relelast}.
Recall that if the elastic material is isotropic, then $\Lambda$ and hence also the stored energy function
$\epsilon$ depends only on the invariants
$\lambda_i$ of
$\gamma^{AB}$, defined with respect to the body metric $\delta_{\Bo}$,
cf. section \ref{sec:relelast}.
Consequently, in view of the discussion above, see in particular section
\ref{sec:stress},  the reduced energy density
$\rho = n \epsilon$ can be viewed as a function $\rho =
\rho(\lambda_i)$.

The invariants $\lambda_i$ are functions of the form
$\lambda_i = \lambda_i(f,\partial f, U, \psi_i, h_{ij};\eta^i,\Omega)$. In the present case, we are using a coordinate system
on $\Bo$ in which the metric $\delta_{\Bo}$ has constant components, and
hence the $\lambda_i$ do not depend on $f$ but only on its derivatives.
We may therefore write
$\rho$ as a functional $\rho = \rho[f,g;\eta,\Omega]$, where the
symbol $g$ is used as shorthand for
the gravitational variables $U, \psi_i, h_{ij}$ parametrizing the spacetime
metric $g_{\mu\nu}$.

\section{Analytical setting} \label{sec:analytical}
We now introduce the analytical setting which will be used to construct
solutions to the field equations. Fix a weight $\delta \in (-1,
-\half)$. Further, fix $p > 3$. The parameters $\delta,p$ will be used to
determine the weighted
Sobolev spaces which are used in the implicit function argument.

The system of equations in material form has the unknowns $\phi^i, U, \psi^i
, h_{ij}$. Let
$$
B_1 = W^{2,p}(\Bo) \times W^{2,p}_\delta \times
W^{2,p}_{\delta} \times E^{2,p}_{\delta}  ,
$$
where $E^{2,p}_{\delta}$ is the space of asymptotically Euclidean metrics
introduced in \cite[section 2.3]{ABS},
and let
$$
B_2 = [L^p(\Bo) \times B^{1-1/p,p}(\partial \Bo)] \times L^p_{\delta-2}
\times L^p_{\delta-2} \times L^p_{\delta-2} \,.
$$
Thus, $B_1$ is a Banach manifold and $B_2$ is a Banach space.

The residuals of equations (\ref{eq:redein4-mat}) with boundary condition
(\ref{eq:redein5-mat}), (\ref{eq:redein1-mat}), (\ref{eq:reduced-psi-mat}),
  (\ref{eq:reduced-h-mat}), which depend on the Newton constant $G$ and the
rotation velocity $\Omega$ as parameters
 define a map $\FF : \Re^2 \times B_1 \to B_2$. Thus, $\FF$ has components
 $(\FF_\phi, \FF_U, \FF_\psi, \FF_h)$ corresponding to the components of
 $B_2$, given by
\begin{subequations}\label{eq:Fdef}
\begin{align}
\FF_\phi &= \left [ \FF^{\Bo}_\phi , \FF^{\partial \Bo}_\phi \right ] \,, \\
\intertext{where}
\FF_\phi^{\Bo} &=
D_A (e^{\bU} \sigma_i{}^A) -  e^{\bU}(\epsilon - \frac{\bar
  \sigma_\ell{}^\ell}{\bar n} ) \overline{\partial_i U} \,, \nonumber \\
&\quad \hskip 0.5in
  - \Omega
[ D_A (e^{\bU} \bar z_i{}^A) + 2  e^{\bU} \frac{\overline{\omega_{ij}
  z^j}}{\bar n} +  \bar z \overline{\partial_i U}] \,, \label{eq:FdefphiBo} \\
\FF^{\partial \Bo}_\phi &=
(\bar \sigma_i{}^A - \Omega \bar z_i{}^A) n_A \big{|}_{\partial \Bo} \,,
\label{eq:FdefphidBo} \\
\intertext{and}
\FF_U &= \overline{\Delta_h U} -  4 \pi G \chi_{\Bo}
e^{\bU}( n \bar \epsilon - \bar \sigma_\ell{}^\ell) + \Omega \bar z)
+ e^{4\bU} \overline{\omega_{kl}\omega^{kl}} \,,
\label{eq:FdefU}
\\
\FF_\psi &=
\overline{D^i (e^{4U} \omega_{ij})}
 + \half \overline{D_j (e^{4U} D_i \psi^i)}
-
8 \pi G
\,\chi_{\Bo}e^{\bU} \Omega \bar z_j \,,
\label{eq:Fdefpsi}
\\
\FF_h &=
-\frac{1}{2}\overline{\Delta_h h_{ij}} + Q_{ij}(\bar h,\partial \bar h)
+ 8 \pi
G e^{\bU} (\bar \sigma_{ij}  - \bar h_{ij}\,\bar \sigma_l{}^l
+ \Omega (\bar z_{ij} - \bar h_{ij}
\bar z_l{}^l))\chi_{\Bo} \nonumber \\
&\quad - 2 \overline{D_i U D_j U }
- e^{4\bU} [ \bar h_{ij} \overline{\omega_{kl} \omega^{kl}} - 2
  \overline{\omega_{ik} \omega_j{}^k} ] \,.
\end{align}
\end{subequations}
We now have $\FF = \FF((G, \Omega), (\phi, \bar U, \bar \psi, \bar
h))$.
Write a general element of $B_1$ as $Z$. We will use the implicit
function theorem to construct solutions to $\FF = 0$ for $G,\Omega$ close to
$0 \in \Re^2$.

An essential assumption which allows us to introduce a relaxed configuration
is that there is a reference Euclidean metric
$\hat{\delta}$ on $M = \Re^3_{\Sp}$, and a diffeomorphism $\ii: \Re^3_{\Bo}
\to \Re^3_{\Sp}$. As discussed in section \ref{sec:constitutive},
an Euclidean metric on $\Re^3_{\Bo}$ is defined by
$\delta_{\Bo} = \ii^* \hat{\delta}$. Recall that $\Bo$ is assumed to be a
connected domain with smooth boundary.

From the constitutive conditions, cf. section \ref{sec:constitutive}
we have that
$$
Z_0 = (\ii, 0 , 0, \hat \delta_{ij} \circ \ii)
$$
is a solution to the equation $\FF(0,Z_0) = 0$. In order to apply the
implicit function theorem at $(0, Z_0)$
it is necessary that the Frechet derivative
$D_2 \FF(0, Z_0)$ is an isomorphism. We see that $\FF(0, Z)$ is of the form
\begin{align*}
\FF_\phi (0,Z) &= \left [
D_A (e^{\bU} \sigma_i{}^A) -  e^{\bU}(\epsilon - \frac{\bar
  \sigma_\ell{}^\ell}{\bar n} ) \overline{\partial_i U} , \quad
\bar \sigma_i{}^A  n_A \big{|}_{\partial \Bo}
\right ] \,, \\
\FF_U(0,Z) &= \overline{\Delta_h U}
+ e^{4\bU} \overline{\omega_{kl}\omega^{kl}} \,,
\\
\FF_\psi(0,Z) &=
\overline{D^i (e^{4U} \omega_{ij})}
 + \half \overline{D_j (e^{4U} D_i \psi^i)} \,,
\\
\FF_h(0,Z) &=
-\frac{1}{2}\overline{\Delta_h h_{ij}} + Q_{ij}(\bar h,\partial \bar h)
 - 2 \overline{D_i U D_j U }
- e^{4\bU} [ \bar h_{ij} \overline{\omega_{kl} \omega^{kl}} - 2
  \overline{\omega_{ik} \omega_j{}^k} ] \,.
\end{align*}
It follows from the constitutive conditions stated in section
  \ref{sec:constitutive} that $D_\phi \FF_\phi(0,Z)$ is elliptic.

\subsection{Projected system} \label{sec:projected}
An analysis along the lines of
\cite[section 4.2]{ABS} shows that $D_2 \FF(0, Z_0)$ is of the form
$$
\begin{pmatrix} D_\phi \FF_\phi & D_U \FF_\phi & 0 & D_h \FF_\phi \\
0 & \Delta & 0 & 0 \\
0 & 0 & \half \Delta & 0 \\
0 & 0 & 0 & - \half \Delta
\end{pmatrix} \,,
$$
where the entries are evaluated at $Z_0$.
The diagonal entries are isomorphisms between the weighted spaces given in
the definition of $B_1$ and $B_2$, with the exception of $D_\phi
\FF_\phi$. As in the static case this has a nontrivial kernel and cokernel,
see the discussion in \cite[section 4]{ABS}. The kernel and cokernel can be
identified with the space of Killing fields on $(\Bo, \delta_{\Bo})$.
Therefore, in order to construct
solutions, we will consider the projected system
$$
\BProj \FF = 0 \,,
$$
where $\BProj : B_2 \to B_2$ is a projection operator
which is defined exactly along the lines of \cite[section 4]{ABS}. In
particular, $\BProj$
is defined to act
as the identity in the second to fourth components of $B_2$, while
in the
first component of $B_2$ it acts
as the unique projection along  the cokernel of
$D_\phi \FF_\phi(0,Z_0)$ onto the range of $D_\phi \FF_\phi (0,Z_0)$, which
leaves the boundary data in the first component of $B_2$
unchanged.  We use the the label $\Bo$ to indicate that $\BProj$ operates on
fields on the body and the extended body.  We shall later need to transport
the projection operator to fields on $\Re^3_{\Sp}$.

Let $(b_i, \tau_i)$ denote pairs of elements in
$W^{2,p}(\Bo) \times W^{1-1/p,p}(\partial \Bo)$.
The restriction of
$\BProj$ to the first component of $B_2$,
which we here denote by the same symbol, is defined by setting
$\BProj (b_i,\tau_i) = ({b'}_i , \tau_i)$, satisfying
\begin{equation}\label{eq:equilib}
\int_{\Bo} \xi^i {b'}_i = \int_{\partial \Bo} \xi^i \tau_i \,,
\end{equation}
for all Killing fields $\xi^i$. Pairs $({b'}_i, \tau_i)$ satisfying this
condition are called equilibrated.  As discussed in \cite[section 4]{ABS},
the definition of $\BProj$ implies there is a unique $\eta_i$ of the form
$\eta_i = \alpha_i + \beta_{ij} X^j$, for constants $\alpha_i, \beta_{ij}$
satisfying $\beta_{ij} = - \beta_{ji}$, such that
$$
{b'}_i = b_i - \eta_i \chi_{\Bo} \,.
$$

We further restrict the domain of $\BProj \FF$
to eliminate the kernel of $D_\phi \BProj \FF$.
By assumption, cf. section \ref{sec:constitutive}, $\Bo$ has an axis of
symmetry, which without loss of generality can be identified with the
$X^3$-axis. Fix a point $X_0$ on the axis of symmetry of $\Bo$, i.e. $X_0$ has
coordinates $(0,0,X^3)$ for some $X^3$.
Recall that the kernel of $D_\phi \FF$ consists of the Killing fields of
$(\Bo, \delta_{\Bo})$. A killing field in $\Bo$ is determined by
specifying its value and antisymmetrized derivative at one point.
Following the proof of
\cite[Proposition 4.3]{ABS}, define\footnote{The discussion here corrects
  some typos in the proof of \cite[Proposition 4.3]{ABS}, in particular the
  antisymmetrization in (\ref{eq:killkernel}) corrects the corresponding
  expression in \cite{ABS}.}
$\XX$ to be the submanifold of $B_1$ such that
\begin{equation} \label{eq:killkernel}
(\phi - \ii)(X_0) = 0 \,, \quad \text{ and }
\delta^C{}_i \delta_{C[A} \partial_{B]} (\phi - \ii)^i(X_0) = 0 \,.
\end{equation}
and define $\YY$ to be the range of the projection operator $\BProj$.
An application of the implicit function theorem to the map
$$
\BProj \FF : \XX \to \YY
$$
now gives the following result, analogous to \cite[Proposition 4.3]{ABS}.
\begin{prop} \label{prop:projected:implicit}
Let $\FF : B_1 \to B_2$ be map defined by (\ref{eq:Fdef}) and let $\BProj$ be
  defined as in \cite[section 4.3]{ABS}. Then, for sufficiently small values
  of Newton's constant $G$ and the rotation velocity $\Omega$,
there is a unique
solution $Z = Z(G,\Omega)$, where $Z = (\phi,
  \bU, \bar \psi_i , \overline{h_{ij}})$,
to the reduced, projected equation for a self-gravitating rotating elastic
body given by
\begin{equation}\label{eq:redproj}
\BProj \FF((G,\Omega), Z) = 0 \,,
\end{equation}
which satisfies the condition (\ref{eq:killkernel}).
In particular, for any $\eps > 0$, there are $G  > 0, \Omega > 0$,
such that $Z(G,\Omega)$
satisfies the inequality
\begin{equation}\label{eq:Gbarsmall}
|| \phi - \ii ||_{W^{2,p}(\Bo)}
+ ||\overline{h_{ij}} - \delta_{ij} ||_{W^{2,p}_\delta}
+ ||\bU||_{W^{2,p}_\delta}
+ ||\bar \psi||_{W^{2,p}_\delta} < \eps \,.
\end{equation}
\end{prop}
The proof of proposition \ref{prop:projected:implicit} proceeds along exactly
the same lines as the proof of \cite[Proposition 4.3]{ABS} and is left to the
reader.

\section{Equilibration} \label{sec:equilibration}
Arguing along the lines of \cite[section 5]{ABS}, we have the following
corollary to Proposition \ref{prop:projected:implicit}.
\begin{cor} \label{cor:GOmegasmall}
For any $\eps > 0$, there are  $G > 0, \Omega > 0$ such
that the inequality
\begin{equation}\label{eq:Gsmall}
|| \phi - \ii ||_{W^{2,p}(\Bo)} + ||h_{ij} - \delta_{ij}
   ||_{W^{2,p}_\delta} + ||U||_{W^{2,p}_\delta} + ||\psi||_{W^{2,p}_\delta}
< \eps
\end{equation}
holds.
\end{cor}

\subsection{Eulerian form of the projected equations} \label{sec:eulerian}
Let $\fProj$ be the Eulerian form of the projection operator, defined as in
\cite[section 5.1]{ABS} by
$$
\fProj (n \cdot (b \circ f)) = n  ( \BProj b ) \circ f \,.
$$
Moving to the Eulerian
form of the projected system, we find that we have constructed, for small
$G,\Omega$ a solution $(\phi, U, \psi_i, h_{ij})$ of the following set of
projected equations, which it is convenient to write in terms of the stress
energy components $\tau, \tau_i, \tau_{ij}$.
\begin{subequations}\label{eq:proj-euler}
\begin{align}
\fProjN  ( D^j \tau_{ij} - 2\, \omega_{ij} \tau^j + & (D_i U) (e^{-4 U} \tau +
\tau_k{}^k)  )  = 0 \,, \label{eq:proj-euler-elast} \\
\tau_{ij} n^j|_{\partial f^{-1}(\Bo)}&=0 \,, \label{eq:proj-elast-bound} \\
\Delta_h U =  4 \pi G \,\chi_{f^{-1}(\Bo)}(e^{-4 U} \tau & + \tau{}_k{}^k)
- e^{4U} \omega_{kl}\omega^{kl} \,, \label{eq:proj-euler-U} \\
D^i(e^{4U} \omega_{ij})+\frac{1}{2}D_j(e^{4U}D^i \psi_i)
&=  8 \pi G \,\chi_{f^{-1}(\Bo)}\tau_j \,, \label{eq:proj-euler-psi}  \\
G_{ij} - D_{(i} V_{j)} + \frac{1}{2} h_{ij} D_l V^l &=  8 \pi G
(\chi_{f^{-1}(\Bo)}\tau_{ij} + \Theta_{ij} + \Omega_{ij}) \,.
\label{eq:proj-euler-ein}
\end{align}
\end{subequations}
Let $Y = (f^A, U, \psi_i, h_{ij})$ be the Eulerian form of the
solution to the projected form of the material field
equations constructed in section \ref{sec:projected}. From proposition
\ref{prop:projected:implicit}, the solution is unique. For the purposes here,
we shall need to make the uniqueness property somewhat more explicit. An
analysis of the proof of proposition \ref{prop:projected:implicit} proves the
following corollary.
\begin{cor}\label{cor:proj:unique}
Let the body domain $\Bo$ with metric $\delta_{\Bo}$ be given, with the
corresponding background metric $\hat \delta$ on $M = \Re^3_\Sp$, and fix a
point $X_0$ in $\Bo$ and a vector field $\eta^A$ on $\Bo$. Then the Eulerian
form $Y = (f^A, U, \psi, h)$ of the solution to the reduced projected system
for a self-gravitating, rotating, elastic body defines a function of the form
$$
Y = Y(G,\Omega;[\Bo,\delta_\Bo,\hat \delta, X_0,\eta]) \,.
$$
\end{cor}

\subsection{Equivariance} \label{sec:equivariance}
We now analyze some of the consequences of the constitutive conditions
  imposed in section \ref{sec:constitutive}. Recall that in particular, in
  view of frame indifference and homogeneity, and the discussion in section
  \ref{sec:constitutive}, the reduced stored energy function $\rho$
is of the form
$$
\rho = \rho[f, g; \eta,\Omega] \,,
$$
where $f^A$ is the configuration, $g$ is used as shorthand for the fields
$U,\psi_i,h_{ij}$ on $M$ parametrizing the
spacetime metric $g_{\mu\nu}$, and $\eta^i$
is the axial vector field on $M$ specified in
section \ref{sec:constitutive}.
Let $\sigma$ be a spatial diffeomorphism, i.e. $t \circ \sigma = t$.
Then by frame indifference (i.e. general covariance) we have
\begin{equation}\label{eq:iso-space}
\rho[f \circ \sigma; \sigma^* g; \sigma_* \eta , \Omega] =
\rho[f,g;\eta,\Omega] \circ \sigma \,.
\end{equation}
Further, as a consequence of the isotropy
of the elastic body, for any
isometry $\Sigma$ of $(\Bo, \delta_{\Bo})$, we have
\begin{equation}\label{eq:iso}
\rho[\Sigma \circ f, g; \eta, \Omega]
= \rho[f,g;\eta, \Omega] \,.
\end{equation}
The transformation properties stated in (\ref{eq:iso-space}) and
(\ref{eq:iso}) give the following Lemma.
\begin{lemma}\label{lem:equi} Let $(f^A,U,\psi_i,h_{ij})$ be as in corollary
  \ref{cor:proj:unique}.
Let $\Sigma$ be a diffeomorphism of $\Bo$ leaving
the data $(X_0, \delta_{AB}, \eta^A)$ invariant.
Then the diffeomorphism $\sigma$ of $M$ defined by requiring that
$\ii \circ \Sigma = \sigma \circ \ii$ on all of $\Re^3_\Bo$ is an isometry
in the sense that it leaves all of $(f^A,U,\psi_i,h_{ij},\eta^i)$ invariant.
\end{lemma}
\begin{proof}
First note that $\sigma$ is by construction an isometry of the flat
background metric $\hat \delta$ entering the projected, harmonically-reduced
field equations and that
$(\sigma_*\eta)^i = \eta^i$ trivially from the construction
of $\eta^i$. Using these facts together with the equivariance property
expressed in (\ref{eq:iso-space}) and
(\ref{eq:iso}) we have that
\begin{equation}\label{eq:order}
((\Sigma^{-1}\circ f \circ \sigma)^A,\sigma^*U,(\sigma^*\psi)_i,
(\sigma^*h)_{ij})
\end{equation}
is a solution with the same data.
By the uniqueness property made explicit in corollary \ref{cor:proj:unique},
we then have
\begin{equation} \label{eq:isometry}
((\Sigma^{-1}\circ f \circ \sigma)^A,\sigma^*U,(\sigma^*\psi)_i,
(\sigma^*h)_{ij}) = (f^A, U, \psi_i, h_{ij} ) \,.
\end{equation}
\end{proof}
If $\sigma$ is as in lemma \ref{lem:equi}, then we also have
\begin{equation}\label{eq:sym}
\sigma^\ast \tau = \tau\,, \quad
(\sigma^\ast \tau)_i = \tau_i\,,
\quad
(\sigma^\ast
\tau)_{ij} = \tau_{ij} \,.
\end{equation}

Lemma \ref{lem:equi} has the following corollary which will play an important
in the proof of orthogonal transitivity, see section \ref{sec:orth-trans}
below.
\begin{cor}\label{cor:reflect} Let $(f^A, U, \psi_i, h_{ij})$ be as in
  corollary \ref{cor:proj:unique}.
Let $\Sigma$ be an isometry of $(\Bo,
\delta_\Bo)$ such that $\Sigma (X_0) = X_0$ and
$(\Sigma_* \eta)^A = - \eta^A$, and let $\sigma$ be a
diffeomorphism of $M$ such that $\ii \circ \Sigma = \sigma \circ \ii$ on
all of $\Re^3_\Bo$.
Then $\sigma$ is an isometry of $h_{ij}$ and we have
\begin{equation}\label{eq:reflect}
(\Sigma^{-1} \circ f^A \circ \sigma , \sigma^* U, (\sigma^* \psi)_i , (\sigma^*
  h)_{ij} ) = (f^A, U, - \psi_i , h_{ij}) \,.
\end{equation}
\end{cor}
\begin{proof} The
  transformation $\psi_i \to - \psi_i$, $\eta^A \to -\eta^A$ leaves $H^{AB}$
and  hence all the field equations invariant. Therefore it maps a solution to
another solution. By uniqueness it follows that the solution with data
$\Bo, \delta_\Bo, \hat \delta, -\eta^A, G, \Omega$ is given by
$(f^A, U, - \psi_i , h_{ij})$. The result follows.
\end{proof}

Recall that the reference state is axially symmetric, i.e. $\eta^A$ is an axial Killing vector
in Euclidean space leaving $\Bo$ invariant.
Denoting by $\Sigma$ the flow of $\eta^A$
and correspondingly using $\sigma$ to denote the flow of $\eta^i$,
we have the following
infinitesimal version of of Lemma \ref{lem:equi}.
\begin{lemma}\label{lem:axial}
Assume that $\Bo$ is axially symmetric with axial Killing field $\eta^A$, as
discussed in section \ref{sec:constitutive}. Then
\begin{equation}\label{eq:relaxi}
f^A{}_{,i} (x)\, \eta^i (x) = \eta^A (f(x)) \,,
\end{equation}
i.e.
\begin{equation}\label{eq:etapullback}
\eta^i \partial_i = f^* (\eta^A \partial_A) \,.
\end{equation}
and
\begin{subequations}\label{eq:kilUpsih1}
\begin{alignat}{1}
\label{eq:kilU1} \Lie_\eta U & = 0 \,,\\
\label{eq:kilpsi1} \Lie_\eta \psi_i & = 0 \,,\\
\label{eq:kilh1} \Lie_\eta h_{ij} & = 0 \,.
\end{alignat}
\end{subequations}
\end{lemma}
By the antisymmetry of $\omega_{ij}$ we have, using (\ref{kilpsi2}), that
\begin{equation}\label{eq:etaomega}
\Lie_\eta(e^{-2U} \alpha) = 0 \,.
\end{equation}
Furthermore, from (\ref{eq:sym}) applied to the flow of $\eta^i$, we
infer that
\begin{equation}\label{eq:sym1}
\Lie_\eta \tau = 0\,, \quad \Lie_\eta \tau_i = 0 \,,\quad \Lie_\eta
\tau_{ij} = 0 \,.
\end{equation}

\subsection{Divergence identities} \label{sec:divergence}
Now turn back to Eq.(\ref{master1}). Taking the divergence of this equation
and using (\ref{eq:kilU1},\ref{eq:kilh1},\ref{eq:etaomega},\ref{eq:sym1}) and
(\ref{eq:proj-euler-elast}), gives
\begin{align}
0 &= (1-\Omega e^{-2U} \alpha) D^i \tau_i
- \Omega D^i (e^{-2U} \alpha) \tau_i \nonumber \\
&\quad
+ \Omega \eta^j (\fIdN - \fProjN)
[ D^i \tau_{ij} - 2 \omega_{ji} \tau^i + (D_j U)(e^{-4U} \tau + \tau_k{}^k)] \\
&\quad + 2 \Omega \eta^j \omega_{ji} \tau^i \nonumber \\
\intertext{use (\ref{kilpsi})}
&= (1-\Omega e^{-2U} \alpha) D^i \tau_i \nonumber \\
&\quad + \Omega \eta^j (\fIdN - \fProjN)
[ D^i \tau_{ij} - 2 \omega_{ji} \tau^i + (D_j U)(e^{-4U} \tau + \tau_k{}^k)]
\chi_{f^{-1}(\Bo)} \,. \label{eq:Ditaui}
\end{align}
It also follows directly from (\ref{master1}) and the fact that $\eta^i$
is parallel to the boundary of $f^{-1} (\Bo)$
that the boundary condition
\begin{equation}\label{eq:bound1-late}
\tau_i n^i|_{f^{-1}(\partial \Bo)}=0
\end{equation}
holds.
Let $W = e^{4U} D^i \psi_i$. The first term in the left hand side of
(\ref{eq:proj-euler-psi}) is the divergence of a 2-form, and therefore its
divergence is zero. Hence, taking the divergence of both sides of
(\ref{eq:proj-euler-psi}), and using the fact that (\ref{eq:bound1-late})
holds for the case of an axisymmetric body, gives
the identity
\begin{equation}\label{eq:DeltaW}
\Delta_h W = 16\pi G \chi_{f^{-1}(\Bo)} D^i \tau_i \,.
\end{equation}
Equation (\ref{eq:Ditaui}) gives the form of the right hand side in
(\ref{eq:DeltaW}).
Let
\begin{equation}\label{eq:Ldef}
L V_i = \Delta_h V_i + R_i{}^k V_k \,,
\end{equation}
and note
$$
D^j (D_{(i} V_{j)} - \half h_{ij} D_k V^k) = \half L V_i \,.
$$
Using (\ref{id1}) and (\ref{id2}) we find after taking the divergence of both
sides of (\ref{eq:proj-euler-ein}), when $G \neq 0$, that
\begin{align*}
LV_i &= - 16\pi G \chi_{f^{-1}(\Bo)} [ D^j \tau_{ij} + (D_i U)(e^{-4U} \tau +
  \tau_k{}^k)] + 4 \omega_{ik} D^j(e^{4U} \omega_j{}^k) \\
\intertext{use (\ref{eq:proj-euler-psi}) and (\ref{eq:proj-euler-elast})}
&= - 16\pi G  [ D^j \tau_{ij} + (D_i U)(e^{-4U} \tau +
  \tau_k{}^k) ] \chi_{f^{-1}(\Bo)}
+ 4 \omega_{ij} [ 8 \pi G \chi_{f^{-1}(\Bo)}\tau^j - \half D^j W] \\
&\quad + 16\pi G  \fProjN [ D^j \tau_{ij} + (D_i U)(e^{-4U} \tau +
  \tau_k{}^k)  - 2 \omega_{ij} \tau^j ] \chi_{f^{-1}(\Bo)} \\
&=   - 16\pi G  (\fIdN - \fProjN )[ D^j \tau_{ij} + (D_i U)(e^{-4U} \tau +
  \tau_k{}^k)  - 2 \omega_{ij} \tau^j ] \chi_{f^{-1}(\Bo)} \\
&\quad - 2 \omega_{ij} D^j W \,.
\end{align*}
Let
$$
\ZZ_i = - 16\pi G [ D^j \tau_{ij} + (D_i U)(e^{-4U} \tau +
  \tau_k{}^k)  - 2 \omega_{ij} \tau^j ] \,.
$$
Then we have the following system of equations for $W, V_i$,
\begin{align*}
(1-\Omega e^{-2U}\alpha)\Delta W &=  \Omega  \eta^j (\fIdN - \fProjN) \ZZ_j
  \chi_{f^{-1}(\Bo)} \,, \\
L V_i  &= (\fIdN - \fProjN) \ZZ_i \chi_{f^{-1}(\Bo)} - 2 \omega_{ij} D^j W \,.
\end{align*}
Arguing as in the proof of \cite[Lemma 5.7]{ABS}, we have that
$$
(\fIdN - \fProjN) \ZZ_i \chi_{f^{-1}(\Bo)} = n (\zeta_i \circ f)
\chi_{f^{-1}(\Bo)} \,,
$$
for some $\zeta_i$ which is a Killing field in $\Bo$.
Hence we have the equations
\begin{subequations} \label{eq:WVeq}
\begin{align}
(1-\Omega e^{-2U}\alpha)\Delta W &=  \Omega  \eta^j n (\zeta_j \circ f)
  \chi_{f^{-1}(\Bo)} \,, \label{eq:Weq} \\
L V_i  &= n (\zeta_i \circ f) \chi_{f^{-1}(\Bo)} - 2 \omega_{ij} D^j W \,.
\label{eq:Veq}
\end{align}
\end{subequations}

\subsection{Main theorem} \label{sec:mainthm}
We are now able to prove the following
\begin{thm}\label{thm:mainthm}
For sufficiently small values of $G,\Omega$, with $G$ non-zero,
the solution to the reduced,
projected system of equations for a stationary, rotating, elastic,
self-gravitating body
(\ref{eq:redproj}), is a solution to the full system of equations
(\ref{ein123}) for a stationary, rotating elastic, self-gravitating body,
together with the integrability conditions of section
\ref{sec:integrability}.
In particular, this solution corresponds to a pair
$(f^A, g_{\mu\nu})$,
which solves the full Einstein equations $G_{\mu\nu} = 8 \pi G T_{\mu\nu}$.
\end{thm}
\begin{proof}
Using the estimate of Corollary
\ref{cor:GOmegasmall}, and the multiplication properties of the weighted
Sobolev spaces, cf. \cite[section 2.3]{ABS}, one checks that
$$
\omega_{ij} \in W^{1,p}_{\delta -1} \,, \quad
\Omega_{ij} \in W^{1,p}_{2\delta-2} \,, \quad
\Theta_{ij} \in W^{1,p}_{2\delta-2} \,,
$$
with corresponding estimates.
Hence we find, using equation (\ref{eq:proj-euler-ein}) for $h_{ij}$,
in the equivalent
form (\ref{eq:reduced-h}),
 equation (\ref{eq:proj-euler-psi}) for $\psi_i$,
making use of
equation (\ref{eq:psi-ellip}) to express it in a form suitable for estimates,
as well as equation (\ref{eq:proj-euler-U}) for $U$,
that the conclusion of \cite[Lemma 5.2]{ABS} for
$h_{ij}$ holds also in the present case, namely
$$
h_{ij} = \delta_{ij} + \frac{\gamma_{ij}}{r}  + h_{(2)\, ij} \,,
$$
for constants $\gamma_{ij}$, with $h_{(2)\, ij} \in W^{2,p}_{2\delta}$.
For sufficiently small $G,\Omega$ we have the estimate
$$
||h_{(2)\ ij} ||_{W^{2,p}_{2\delta}} + ||\gamma|| \leq C
( ||h_{ij} - \delta_{ij} ||_{W^{2,p}_{\delta}}
+ ||\phi - \ii||_{W^{2,p}(\Bo)}
) \,.
$$
For brevity, we shall in the following write estimates of the above form
using $||Z - Z_0||_{B_1}$ where the norm refers to that induced from the Banach
spaces using in defining the space $B_1$, cf. section
\ref{sec:analytical}. We shall further write inequalities of the form
$a \leq C b$ where $C$ is a constant which is uniformly bounded for small $G,
\Omega$ as $a \oleq b$.

Given this result about the asymptotics of $h_{ij}$, the conclusion of
\cite[Lemma 5.4]{ABS} concerning $V_i$ holds, and hence also the partial
integration result \cite[Lemma 5.5]{ABS} and the estimate of
\cite[Lemma 5.6]{ABS}.
Now define the operator $\QQ: L^p_{\delta-3}(\Re^3_{\Sp}) \to \Re^6$ as in
\cite[section 5.2]{ABS}.
Given a basis
$\{ \xi_{(\kappa)} \}_{\kappa = 1}^6$ for the space of Killing fields,
we set
$$
\QQ_{\kappa} (z_i) = \int_{\Re^3_{\Sp}} (\xi^i_{(\kappa)} \circ f )z_i d\mu_h , \quad
\kappa = 1,\dots,6 \,.
$$
Since $W = e^{4 U} D^i \psi_i$, the term $\omega_{ij} D^j W$ in (\ref{eq:Veq}) satisfies
$\omega_{ij} D^j W \in L^p_{2\delta - 3}$ and we have the
estimate
\begin{equation} \label{eq:omDW-est}
||\omega_{ij} D^j W ||_{L^p_{\delta-3}(\Re^3_{\Sp})} \oleq
||Z - Z_0||_{B_1} ||n ( \zeta \circ f) ||_{L^p(f^{-1}(\Bo))} \,.
\end{equation}
Recall that from the construction of $\QQ$ we have for small $G, \Omega$, the equivalence of norms
\begin{equation}\label{eq:normequiv}
|| \QQ n (\zeta \circ f) \chi_{f^{-1}(\Bo)} ||_{\Re^6} \oleq
||\zeta||_{\Re^6} \oleq ||\QQ n (\zeta \circ f) \chi_{f^{-1}(\Bo)}||_{\Re^6}
\,,
\end{equation}
where if $\zeta^i = \alpha^i + \beta^i{}_j x^j$, $||\zeta||_{\Re^6}$
is defined by
$$
||\zeta||_{\Re^6}^2  = \sum_i (\alpha^i)^2 + \sum_{i< j} (\beta^i{}_j)^2 \,.
$$
Due to the properties of $\QQ$, the analogue of (\ref{eq:normequiv}) holds
also for $||n (\zeta \circ f)||_{L^p(f^{-1}(\Bo))}$.
Applying $\QQ$ to both sides of (\ref{eq:Veq}), we have using
(\ref{eq:omDW-est}) and (\ref{eq:normequiv}),
\begin{align*}
|| \zeta ||_{\Re^6} &\oleq ||\QQ LV||_{\Re^6}
+ || \QQ \omega DW ||_{\Re^6} \\
&\oleq ||\QQ LV||_{\Re^6} + ||Z - Z_0||_{B_1} ||\zeta||_{\Re^6} \,,
\end{align*}
and hence
\begin{equation}\label{eq:zetaQLV}
||\zeta||_{\Re^6} \oleq ||\QQ LV||_{\Re^6} \,.
\end{equation}
Recall that for $G, \Omega$ sufficiently small, we also have due to Corollary
\ref{cor:GOmegasmall} that $||Z - Z_0||_{B_1}$ small.
We now have the chain of inequalities for $G, \Omega$ sufficiently small,
\begin{align*}
||V||_{W^{2,p}_{\delta-1}} &\oleq ||LV||_{W^{2,p}_{\delta-3}} \\
&\oleq || n (\zeta \circ f)||_{L^p(f^{-1}(\Bo))}
+ ||\omega DW ||_{L^p_{\delta -3}} \\
&\oleq ||\zeta||_{\Re^6} + ||Z - Z_0||_{B_1} ||\zeta||_{\Re^6} \\
&\oleq ||\zeta||_{\Re^6} \\
\intertext{use (\ref{eq:zetaQLV})}
&\oleq ||\QQ LV||_{\Re^6} \,.
\end{align*}
By the inequality proved in \cite[Proposition 5.8]{ABS}
we have
$$
||\QQ LV||_{\Re^6} \oleq || Z - Z_0||_{B_1} ||V||_{W^{2,p}_{\delta-1}} \,,
$$
which together with the above gives
\begin{equation} \label{eq:Vsmall}
||V||_{W^{2,p}_{\delta-1}} \oleq ||Z - Z_0||_{B_1} ||V||_{W^{2,p}_{\delta-1}}
\,.
\end{equation}
By choosing $G, \Omega$ sufficiently small, we can make $||Z - Z_0||_{B_1}$
small enough so that (\ref{eq:Vsmall}) gives the inequality
$$
||V||_{W^{2,p}_{\delta-1}} \leq \half ||V||_{W^{2,p}_{\delta-1}} \,,
$$
which implies
$$
V = 0 \,.
$$
Due to the vanishing of $V$, it follows from (\ref{eq:zetaQLV}) that  also
$\zeta = 0$, and hence we have
$$
W = 0 \,.
$$
This means that the solution of the projected system of equations
(\ref{eq:proj-euler}) is actually a solution to the full system of field
equations (\ref{ein123}) for the rotating elastic body, together with the
integrability conditions discussed in section \ref{sec:integrability}.

It remains to demonstrate that the solution $(f^A, U, \psi_i, h_{ij})$ to
(\ref{ein123}) constructed in this proof
corresponds to a Lorentzian spacetime $(\MM,g_{\mu\nu})$ solving the
Einstein equations for the elastic body.
The solution we have found yields via (\ref{metric}) a Lorentz metric
$g_{\mu\nu}$ at some time $t_0$ together with its vanishing first
and second
time derivatives
at $t_0$, as well as a configuration  $f^A$ together with its non-vanishing
first time derivative at $t_0$. These solve
the Einstein equations at $t_0$. We extend
the spacetime metric off $t_0$ by requiring it to be $t$-independent
and $f^A$ by requiring it to satisfy (\ref{eq:hel}) for all times.
This constructs a spacetime $(\MM, g_{\mu\nu})$
which is axisymmetric and stationary
and a configuration which is axially symmetric and helical. Thus,
by the discussion in section
\ref{sec:material:isometry}, the associated energy momentum tensor
is time independent. This shows that $(\MM, g_{\mu\nu})$ together with the
configuration $f^A$ provide a solution to the full Einstein equations.
\end{proof}
We remark that the solutions we have found are static exactly when
$\Omega = 0$.

\subsection{Orthogonal Transitivity} \label{sec:orth-trans}
Let $(\MM, g_{\mu\nu})$ be a stationary spacetime containing a rotating
elastic body as constructed in Theorem \ref{thm:mainthm}.
We have shown in section \ref{sec:equilibration} that $(\MM, g_{\mu\nu})$
admits a two-parameter, abelian group of isometries, generated by the Killing
fields $\xi^\mu, \eta^\mu$. In fact, since $\eta^\mu$ is the pullback of the
axial vector field acting on the body, the group can be taken to be the
cylinder $\Re \times S^1$. The question arises if this group acts
orthogonally transitively on $\MM$, as is the case for perfect fluids.
Recall
that a group
acts orthogonally
transitively
if the
the  distribution perpendicular to the generators of the
group action is Frobenius integrable.

Define  $\omega_{\mu\nu\lambda} = 3 \xi_{[\mu} \nabla_\nu \xi_{\lambda]}$ and
let ${\omega'}_{\mu\nu\lambda}$ be defined with respect to $\eta_\mu$ in the
analogous manner. Orthogonal transitivity is equivalent to the conditions
\begin{subequations}\label{eq:twosurfacecond}
\begin{align}
\eta_{[\rho} \omega_{\mu\nu\lambda]} &= 0 \,, \label{eq:twosurfacexi} \\
\xi_{[\rho} {\omega'}_{\mu\nu\lambda]} &= 0 \,, \label{eq:twosurfaceeta}
\end{align}
\end{subequations}
see \cite[(2.53)]{beig:schmidt:timeindep}. The spacetimes constructed in this
paper have metrics which fail to be smooth at the boundary of the body
$f^{-1} (\Bo)$.
\begin{prop}\label{prop:OT}
Let $(\MM,g_{\mu\nu})$ be a stationary spacetime containing a rotating
elastic body as in Theorem \ref{thm:mainthm}.
, with stationary and axial
Killing fields $\xi^\mu, \eta^\mu$. Then, if $\Omega > 0$ is sufficiently
small, equation (\ref{eq:twosurfacecond})
holds in $\MM$.
\end{prop}
\begin{proof}
The conditions (\ref{eq:twosurfacecond}) can be restated in the space
manifold $M$ as
\begin{subequations} \label{eq:spacelike-OT}
\begin{align}
e^{2 U} \eta_{[i} \omega_{jk]} &= 0 \,, \label{eq:2.61:BS}\\ e^{-2 U} \eta_{[i}
D_j \eta_{k]} + \alpha \eta_{[i} \omega_{jk]} &= 0 \,.
\label{eq:2.60:BS}
\end{align}
\end{subequations}
Here
$\omega_{ij} = \partial_{[i} \psi_{j]}$ and $\alpha = \xi^\mu \eta_\mu$, as
above. Equations \cite[(2.60),(2.61)]{beig:schmidt:timeindep} are
equivalent to (\ref{eq:spacelike-OT}) but written in terms of a different
representation of the spacetime metric.

By \cite[equations (2.51)-(2.52)]{beig:schmidt:timeindep},
we have
\begin{subequations}\label{eq:2.51-2:BS}
\begin{align}
4 \nabla^\mu ( \eta_{[\rho}{\omega}_{\mu\nu\lambda]}) &=  6 \xi^\mu
R_{\mu[\nu} \xi_\lambda \eta_{\rho]} \,, \label{eq:2.51:BS} \\ 4 \nabla^\mu (
\xi_{[\rho}{\omega'}_{\mu\nu\lambda]}) &= - 6 \eta^\mu R_{\mu[\nu}
\xi_\lambda \eta_{\rho]} \,.\label{eq:2.52:BS}
\end{align}
\end{subequations}
It follows that
\begin{equation}\label{new}
\nabla^\mu ( \eta_{[\rho}{\omega}_{\mu\nu\lambda]} - \Omega
\xi_{[\rho}{\omega'}_{\mu\nu\lambda]}) =  12 \pi G (\xi^\mu + \Omega
\eta^\mu) T_{\mu[\nu} \xi_\lambda \eta_{\rho]} \,.
\end{equation}
Since the velocity vector  $u^\mu = b^{-1} (\xi^\mu + \Omega \eta^\mu)$, see
equation (\ref{eq:u-def}), is an eigenvector of the stress energy tensor,
cf. equation  (\ref{eq:TmunuLam}),  the right hand  side of (\ref{new}) is
zero. The left hand side of (\ref{new}) is the divergence of a 4-form,
i.e. in terms of the exterior derivative and the Hodge dual,
we have an equation of the form $\star\,\dHodge \star \alpha = 0$.
In particular, $\star\alpha$ is a scalar function which is constant,
$\dHodge \star\alpha = 0$.
In the situation under consideration, $\alpha$ vanishes on
the axis $x^1 = x^2 = 0$, and hence it is zero everywhere.
In this argument we made use of the fact that $u^\mu$ is
well defined in all of $f^{-1}(\Bo)$. This holds for sufficiently small
values of $\Omega$, since then the vector field $\xi^\mu + \Omega \eta^\mu$
is timelike in all of $f^{-1} (\Bo)$.  In terms of the space manifold $M$, we
have shown that
\begin{equation}\label{new1}
 \Omega \eta_{[i}D_j  \eta_{k]} - e^{4 U}(1 - \Omega e^{- 2 U}
 \alpha)\eta_{[i}\omega_{jk]} = 0\,.
\end{equation}
This can of course also be checked directly from the three dimensional field
equations.  Note that relation (\ref{new1}) becomes vacuous in the static
case $\Omega = 0$.

By the above argument we have shown that the two equations
(\ref{eq:spacelike-OT}) are linearly dependent if $\Omega \ne 0$.
Thus, in order to show that both equations in
(\ref{eq:spacelike-OT}) hold, it is sufficient to show that
$\eta_{[i} D_j \eta_{k]} = 0$. To see this we argue as follows. It
follows from the axisymmetry of the body that there is a discrete
isometry $\Sigma$ of $(\Bo, \delta_\Bo)$, consisting of reflections
in planes containing the $X^3$ axis, which maps $\eta^A$ to
$-\eta^A$. An explicit choice of $\Sigma$ is given by
$$
\Sigma(X^1,X^2,X^3) = (-X^1, X^2, X^3) \,.
$$
By corollary \ref{cor:reflect}, and the construction of $\eta^i$,
we have that the diffeomorphism $\sigma$ of
$M$ defined by $\Sigma \circ \ii = \sigma \circ \ii$ is an isometry of
$h_{ij}$, which has the property that $(\sigma_* \eta)^i = - \eta^i$.
We can now conclude
that reflections at planes through the $x^3$-axis preserve both $U$ and
$h_{ij}$ and send both $\psi_i$ and $\eta^i$ to their respective
negatives. So in particular these transformations
preserve vectors tangent to
these planes, and since they send $\eta^i$ to $- \eta^i$ and preserve inner
products, $\eta^i$ has to be orthogonal to these planes. Consequently
$\eta^i$ is hypersurface orthogonal, i.e.
\begin{equation}\label{hyper}
\eta_{[i}D_j \eta_{k]} = 0 \,.
\end{equation}
It follows, using (\ref{new1}), that (\ref{eq:spacelike-OT}) holds.
\end{proof}
\begin{remark} \label{rem:stresseigen}
Recall the identity valid for Killing vectors
\begin{equation}\label{identity}
3 D^i(\eta_{[i}D_j\eta_{k]}) = 2 \eta_{[j}R_{k]l}\, \eta^l \,.
\end{equation}
Inserting (\ref{identity}) into (\ref{eq:Gij}), using (\ref{kilpsi2},\ref{kilU}) and finally (\ref{new1}), there results
\begin{equation}\label{shear}
3 D^i[(1 - \Omega e^{- 2 U} \alpha)^{\frac{1}{3}} \eta_{[i}D_j\eta_{k]}] = 16 \pi G (1 - \Omega e^{- 2 U} \alpha)^{\frac{1}{3}}
\eta^i \tau_{i[j}\eta_{k]} \,.
\end{equation}
Thus we have inferred that $\eta^i$ is an eigenvector of the stress tensor.
This latter fact could have also been shown directly from the reflection symmetry
without using the Einstein equations.
\end{remark}
\begin{remark} \label{rem:OT-frob}
In the case of a smooth spacetime, it follows from (\ref{eq:twosurfacecond})
and the Frobenius theorem that the distribution perpendicular to
$\xi^\mu,\eta^\mu$ is integrable, in the sense that there are smooth
2-surfaces in $\MM$ orthogonal to the span of $\xi^\mu, \eta^\mu$.
The spacetimes constructed in Theorem \ref{thm:mainthm} have in this paper
shown to be $W^{2,p}_{\text{loc}}$. Although the spacetimes containing a
rotating body can in fact be shown to be real analytic away from the
boundary of the body, $f^{-1}(\partial \Bo)$, a further analysis is needed
to show that an appropriate version of the Frobenius theorem applies. This
question will be studied in a later paper.
\end{remark}

\appendix

\section{Proof of Lemma \ref{lem:rhotau}} \label{sec:rhotauproof}
We have $\Lambda = e^{3U} \rho$. From
$$
T_{\mu\nu} = 2\frac{\partial \Lambda}{\partial g^{\mu\nu}} - \Lambda
g_{\mu\nu} \,,
$$
we get
$$
\frac{\partial \Lambda}{\partial g^{\mu\nu}} = \half ( T_{\mu\nu} + \Lambda
g_{\mu\nu} ) \,.
$$
Using the form of $g^{\mu\nu}$, cf. (\ref{imetric}),
we have
\begin{align*}
\frac{\partial g^{\mu\nu}}{\partial h^{ij}} \partial_\mu \partial_\nu &=
e^{2U} (\partial_i \partial_j  - \psi_i \psi_j \partial_t^2 ) \,,
\\
\frac{\partial g^{\mu\nu}}{\partial \psi^i} \partial_\mu \partial_\nu
&= e^{2U} ( - 2 \partial_i \partial_t + 2 \psi_i \partial_t^2) \,,
 \\
\frac{\partial g^{\mu\nu}}{\partial U} \partial_\mu \partial_\nu &=
2 g^{\mu\nu} \partial_\mu \partial_\nu + 4 e^{-2U} \partial_t^2 \,.
\end{align*}
Define $\tau, \tau_i, \tau_{ij}$ by
$$
T_{\mu\nu} = \tau (dt + \psi_i dx^i)^2 + 2 \tau_j dx^j (dt + \psi_i dx^i) +
\tau_{ij} dx^i dx^j \,.
$$
Then,
\begin{align*}
T_{ij} &= \tau_{ij} + 2 \tau_{(i} \psi_{j)} + \tau \psi_i \psi_j \,, \\
T_{0i} &= \tau_i + \tau \psi_i \,, \\
T_{00} &= \tau \,, \\
T_\mu{}^\mu &= - e^{-2U} \tau + e^{2U} \tau_\ell{}^\ell \,.
\end{align*}
We calculate
\begin{align*}
e^U  (2 \frac{\partial \rho}{\partial
  h^{ij}} - \rho h_{ij} ) &= e^{-2U} (2 \frac{\partial \Lambda}{\partial
  h^{ij}} - \Lambda h_{ij} )  \\
&= e^{-2U} (2 \frac{\partial \Lambda}{\partial
  g^{\mu\nu}} \frac{\partial g^{\mu\nu}}{\partial h^{ij}} - \Lambda h_{ij} )
\\
  &= e^{-2U} [ ( T_{\mu\nu} + \Lambda g_{\mu\nu} ) \frac{\partial
  g^{\mu\nu}}{\partial h^{ij}}  - \Lambda h_{ij} ) \\
&= T_{ij} - T_{00} \psi_i \psi_j + \Lambda ( g_{ij} - g_{00} \psi_i \psi_j )
  - \Lambda e^{-2U} h_{ij} \\
&= \tau_{ij} \,, \\
e^U \frac{\partial \rho}{\partial \psi^i} &= e^{-2U} \frac{\partial
  \Lambda}{\partial \psi^i} \\
&= e^{-2U} \frac{\partial \Lambda}{\partial g^{\mu\nu}} \frac{\partial
  g^{\mu\nu}}{\partial \psi^i} \\
&= e^{-2U} \half ( T_{\mu\nu} + \Lambda g_{\mu\nu} ) \frac{\partial
  g^{\mu\nu}}{\partial \psi^i} \\
&= - T_{i0} + \psi_i T_{00} - \Lambda g_{i0} + \Lambda \psi_i g_{00} \\
&= - \tau_i \,, \\
e^U (\frac{\partial \rho}{\partial U} + \rho ) &= e^{-2U} (\frac{\partial
  \Lambda}{\partial g^{\mu\nu}} \frac{\partial g^{\mu\nu}}{\partial U} - 2
  \Lambda)  \\
&= e^{-2U} ( \half (T_{\mu\nu} + \Lambda g_{\mu\nu})  \frac{\partial
  g^{\mu\nu}}{\partial U} - 2\Lambda ) \\
&= e^{-2U} ( (T_{\mu\nu} + \Lambda g_{\mu\nu} ) ( g^{\mu\nu} + 2 e^{-2U}
  \delta^\mu{}_0 \delta^\nu{}_0 ) - 2\Lambda ) \\
&= e^{-2U} ( T_\mu{}^\mu + 4 \Lambda + 2 e^{-2U} T_{00} + 2 e^{-2U} \Lambda
  g_{00} - 2\Lambda ) \\
&= e^{-2U} (T_\mu{}^\mu + 2 e^{-2U} T_{00} ) \\
&= e^{-4U} \tau + \tau_\ell{}^\ell \,.
\end{align*}

\subsection*{Acknowledgements} LA and RB thank the Mittag-Leffler Institute,
  Djursholm, Sweden,  
where part of this paper was written, for hospitality and support.

\providecommand{\bysame}{\leavevmode\hbox to3em{\hrulefill}\thinspace}
\providecommand{\MR}{\relax\ifhmode\unskip\space\fi MR }
\providecommand{\MRhref}[2]{%
  \href{http://www.ams.org/mathscinet-getitem?mr=#1}{#2}
}
\providecommand{\href}[2]{#2}

\end{document}